\documentclass[a4paper,fleqn,usenatbib]{mnras}

\usepackage[T1]{fontenc}
\usepackage{ae,aecompl}

\usepackage{graphicx}	% Including figure files
\usepackage{amsmath}	% Advanced maths commands
\usepackage{amssymb}	% Extra maths symbols
\usepackage{subfigure}
\usepackage{multirow}
\usepackage{xcolor}

\newcommand{\pcc}{\,{\rm cm}^{-3}}
\newcommand{\gcc}{\,{\rm g \, cm}^{-3}}
\newcommand{\pcs}{\,{\rm cm}^{-2}}
\newcommand{\um}{\, {\rm \mu m}}
\newcommand{\kel}{\, {\rm K}}

\newcommand{\msun}{\, {\rm M}_\odot}
\newcommand{\nh}{n_{\rm H}}
\newcommand{\Nhcol}{N_{\rm H}}

\newcommand{\pc}{\, {\rm pc}}

\newcommand{\kpc}{\, {\rm kpc}}

\newcommand{\myr}{\, {\rm Myr}}
\newcommand{\kyr}{\, {\rm kyr}}
\newcommand{\ug}{\, {\rm \mu G}}

\newcommand{\kms}{\, {\rm km \, s^{-1}}}

\title[The NEATH survey]{NEATH III: a molecular line survey of a simulated star-forming cloud}

\author[Priestley et al.]{
  F. D. Priestley$^1$\thanks{Email: priestleyf@cardiff.ac.uk}, P. C. Clark$^1$, S. C. O. Glover$^2$, S. E. Ragan$^1$, O. Feh\'{e}r$^1$, L. R. Prole$^3$,
  \newauthor R. S. Klessen$^{2,4}$
\\
$^{1}$School of Physics and Astronomy, Cardiff University, Queen's Buildings, The Parade, Cardiff CF24 3AA, UK \\
$^{2}$Universit\"{a}t Heidelberg, Zentrum f\"{u}r Astronomie, Institut f\"{u}r Theoretische Astrophysik, Albert-Ueberle-Stra{\ss}e 2, D-69120 Heidelberg, Germany\\
$^{3}$Centre for Astrophysics and Space Science Maynooth, Department of Theoretical Physics, Maynooth University, W23 F2H6 Maynooth, Ireland \\
$^{4}$Universit\"{a}t Heidelberg, Interdisziplin\"{a}res Zentrum f\"{u}r Wissenschaftliches Rechnen, Im Neuenheimer Feld 205, D-69120 Heidelberg, Germany\\ 
}

\date{Accepted XXX. Received YYY; in original form ZZZ}

\pubyear{2024}

\begin{document}
\label{firstpage}
\pagerange{\pageref{firstpage}--\pageref{lastpage}}
\maketitle

\begin{abstract}

  We present synthetic line observations of a simulated molecular cloud, utilising a self-consistent treatment of the dynamics and time-dependent chemical evolution. We investigate line emission from the three most common CO isotopologues ($^{12}$CO, $^{13}$CO, C$^{18}$O) and six supposed tracers of dense gas (NH$_3$, HCN, N$_2$H$^+$, HCO$^+$, CS, HNC). Our simulation produces a range of line intensities consistent with that observed in real molecular clouds. The HCN-to-CO intensity ratio is relatively invariant with column density, making HCN (and chemically-similar species such as CS) a poor tracer of high-density material in the cloud. The ratio of N$_2$H$^+$ to HCN or CO, on the other hand, is highly selective of regions with densities above $10^{22} \pcs$, and the N$_2$H$^+$ line is a very good tracer of the dynamics of high volume density ($>10^4 \pcc$) material. Focusing on cores formed within the simulated cloud, we find good agreement with the line intensities of an observational sample of prestellar cores, including reproducing observed CS line intensities with an undepleted elemental abundance of sulphur. However, agreement between cores formed in the simulation, and models of isolated cores which have otherwise-comparable properties, is poor. The formation from and interaction with the large-scale environment has a significant impact on the line emission properties of the cores, making isolated models unsuitable for interpreting observational data.

\end{abstract}

\section{Introduction}

Star formation occurs in molecular clouds, where high gas densities and strong shielding from the external ultraviolet (UV) radiation field result in most hydrogen existing in the form of H$_2$ \citep{bergin2007}. Temperatures in molecular clouds are too low to excite any significant line emission from H$_2$, making the bulk of the star-forming gas essentially invisible. Studies of star formation therefore rely on indirect tracers of the gas mass, such as far-infrared (far-IR) thermal emission from dust grains \citep[e.g.][]{konyves2015} or extinction measurements of background stars \citep{zucker2021}. Of particular utility are rotational emission lines from molecular species at millimetre wavelengths. Line emission provides information on the line-of-sight velocity of the gas, crucial for assessing its dynamics and stability, and the large number of lines detectable with modern observational facilities \citep[e.g.][]{pety2017,kauffmann2017,tafalla2021} can be used to estimate properties such as volume density and temperature, exploiting the fact that the excitation of each line has a unique response to the local physical conditions \citep{shirley2015}.

Analysis of molecular line data is complicated by the fact that the observed emission depends on a complex combination of the physical structure of the cloud, radiative transfer effects, and the chemical composition of the gas. Disentangling all of these factors to accurately reconstruct detailed cloud properties from observational data is not practically feasible. Many studies have therefore investigated forward modelling the problem and producing synthetic line observations of simulated clouds, which can be used to help inform a physical interpretation of real observational data. To date, most of these studies have focused on lines from important coolant species, in particular CO \citep[e.g.][]{penaloza2017,seifried2017,penaloza2018,clarke2018,clark2019}, as its chemical evolution is {often} followed self-consistently in modern hydrodynamical simulations. However, due to its high abundance and correspondingly high line optical depths, CO is a poor tracer of the high-density material where star formation actually occurs in molecular clouds \citep{clark2019,priestley2023a}. {Isotopologues of CO, such as C$^{18}$O, have much lower optical depths and hence are better tracers of dense gas, but they are also faint and hard to observe in distant clouds or extragalactic systems.}

Studies investigating lines from rarer molecules such as HCN and N$_2$H$^+$, which can probe denser gas more effectively, generally assume that the molecular abundances are constant, or can be {specified} as a function of the local physical properties \citep[e.g.][]{offner2008,smith2012,smith2013,chira2014,jones2023,jensen2023}. This approach ignores the drastic variations in abundance driven by differing evolutionary histories, which become particularly important at these high densities \citep{priestley2023c}. Obtaining reliable line emission properties from simulated molecular clouds requires the physical and chemical evolution of the clouds to be treated self-consistently.

In this paper, we achieve this by performing radiative transfer simulations of a cloud for which the molecular composition has been obtained from a full time-dependent chemical network. The abundances at each point in the cloud are thus consistent with the physical evolution of the material that ended up at that point, making the resulting line emission data genuinely comparable to that from real molecular clouds with their own complex formation histories. We use this to assess the degree of correspondance between our simulated cloud and reality, the limitations of smaller-scale models in interpreting line emission, and the accuracy of commonly-used observational diagnostics of the physical structure of clouds.

\section{Method}

\begin{table}
  \centering
  \caption{Elemental abundances used in the chemical modelling.}
  \begin{tabular}{ccccc}
    \hline
    Element & Abundance & & Element & Abundance \\
    \hline
    C & $1.4 \times 10^{-4}$ & & S & $1.2 \times 10^{-5}$ \\
    N & $7.6 \times 10^{-5}$ & & Si & $1.5 \times 10^{-7}$ \\
    O & $3.2 \times 10^{-4}$ & & Mg & $1.4 \times 10^{-7}$ \\
    \hline
  \end{tabular}
  \label{tab:abun}
\end{table}

\begin{table}
  \centering
  \caption{Molecules investigated, collisional partners, and sources of collisional rate data.}
  \begin{tabular}{ccc}
    \hline
    Molecule & Partners & Reference \\
    \hline
    $^{12}$CO & p-H$_2$, o-H$_2$ & \citet{yang2010} \\
    $^{13}$CO & p-H$_2$, o-H$_2$ & \citet{yang2010} \\
    C$^{18}$O & p-H$_2$, o-H$_2$ & \citet{yang2010} \\
    p-NH$_3$ & p-H$_2$ & \citet{loreau2023} \\
    \quad \\
    \multirow{2}{*}{HCN} & \multirow{2}{*}{p-H$_2$, e$^-$} & \citet{faure2007} \\
    & & \citet{magalhaes2018} \\
    \quad \\
    N$_2$H$^+$ & p-H$_2$ & \citet{lique2015} \\
    HCO$^+$ & p-H$_2$, o-H$_2$ & \citet{denis2020} \\
    CS & H$_2$ & \citet{lique2006} \\
    HNC & H$_2$ & \citet{dumouchel2010} \\
    \hline
  \end{tabular}
  \label{tab:moldata}
\end{table}

\subsection{Hydrodynamical simulation}

We simulate the formation and subsequent evolution of a molecular cloud via the collision of two spherical, initially-atomic gas clouds. The simulations are performed using the magnetohydrodynamic (MHD) moving-mesh code {\sc arepo} \citep{springel2010,pakmor2011}, modified to properly capture the thermodynamics of the gas and dust \citep{glover2007,glover2012}. This includes a simplified chemical network for H$_2$ and CO ({based on \citealt{gong2017}, with some additions and modifications described in \citealt{hunter2023}}), and a self-consistent treatment of shielding from the background UV radiation field \citep{clark2012}.

The initial conditions of the simulation are two spherical $10^4 \msun$ clouds with radii $R = 19 \pc$, giving an initial volume density of hydrogen nuclei $\nh = 10 \pcc$. The gas and dust temperatures are initialised to $300 \kel$ and $15 \kel$ respectively. The clouds are displaced in the $\pm x$ direction by $R$, so they are just in contact at the outset, and given a bulk motion along the $x$-axis of $\mp 7 \kms$. Each cloud also has a virialised turbulent velocity field with {3D} velocity dispersion $\sigma = 0.95 \kms$, and a $3 \ug$ magnetic field is present parallel to the collision axis. In collapsing high-density regions of the simulation, sink particles are introduced (following \citealt{tress2020}; see also \citealt{prole2022}), with a threshold density of $2 \times 10^{-16} \gcc$ and a formation radius of $9 \times 10^{-4} \pc$. The simulation is run for $5.53 \myr$, at which point sink particles have accreted $102 \msun$ of material ($0.5 \%$ of the original cloud masses); beyond this point, feedback effects from newly-formed stars (which are not modelled) are likely to become important \citep{whitworth1979,walch2012}.

As the mesh cells can be created and destroyed, and so do not correspond to coherent parcels of gas, we follow the physical evolution of representative gas parcels using Monte Carlo tracer particles \citep{genel2013}, which provide the input for the chemical modelling described below. Global properties, shared between the MHD and chemical models, are the UV radiation field strength of $1.7$ times the \citet{habing1968} field, the cosmic ray ionisation rate\footnote{{Our simulations do not include attenuation of cosmic rays, so the ionisation rate is constant with density. The observed rate in high-density regions of molecular clouds \citep{pineda2024,redaelli2024} is typically somewhat lower than the value we have chosen (representative of the diffuse interstellar medium; \citealt{indriolo2015}), which may be an overestimate for the corresponding regions in the simulation.}} of $10^{-16} \, {\rm s^{-1}}$ {per H atom}, the dust-to-gas mass ratio of $0.01$, and the elemental abundances of carbon ($1.4 \times 10^{-4}$) and oxygen ($3.2 \times 10^{-4}$) from \citet{sembach2000}. The `metal' abundance in the MHD simulation, representing heavier elements such as silicon, is set to $10^{-7}$, corresponding to the high levels of depletion found in the dense interstellar medium \citep{jenkins2009}. 

\subsection{Chemical modelling}

The tracer particles in the MHD simulation record the density, gas temperature, and effective shielding column densities at intervals of $44 \kyr$. These particle histories are used to model the chemical evolution under the NEATH framework\footnote{https://fpriestley.github.io/neath/} \citep{priestley2023c}, which is designed so that the returned H$_2$ and CO abundances are consistent with those of the underlying MHD chemical network. We use a modified version of the time-dependent gas-grain code {\sc uclchem} \citep{holdship2017} to evolve the UMIST12 \citep{mcelroy2013} reaction network, with the same UV and cosmic ray values as the MHD simulation. The assumed elemental abundances are listed in Table \ref{tab:abun}. We take carbon, nitrogen and oxygen abundances from \citet{sembach2000}, and deplete silicon and magnesium by factors of $100$ from their warm neutral medium (WNM) values, again representing depletion into refractory dust grains. Unlike most {astrochemical models of star-forming clouds and cores}, we do not deplete sulphur from its WNM value, which is close to the undepleted Solar value; we return to this point in Section \ref{sec:sulphur}.

We follow the chemical evolution of $10^5$ tracer particles chosen from the central $16.2 \pc$ of the computational domain at the simulation's end, which contains virtually all of the molecular material. Particles are selected randomly to evenly sample densities between $10-10^6 \pcc$, the highest density for which the chemical evolution is converged \citep{priestley2023c}. Lower-density material has a negligible molecular content, and higher-density material makes up a negligible fraction of the cloud mass. We confirm that this number and selection of particles is sufficient for our purposes in Appendix \ref{sec:restest}.

\subsection{Radiative transfer}

We perform radiative transfer modelling of our simulated cloud using {\sc radmc3d} \citep{dullemond2012}. The unstructured Voronoi mesh of the MHD simulation is interpolated onto a cubic adaptive mesh, with side length $32.4 \pc$ and a base resoluton of $20^3$. Any cell which contains more than one Voronoi sampling point is refined into eight subcells, and this process is repeated until all cells hold at most one sampling point each. Cells are then assigned physical properties (density, velocity, gas and dust temperatures) from the nearest Voronoi sampling point, and molecular abundances from the nearest post-processed tracer particle. We assess the accuracy of the interpolation in Appendix \ref{sec:restest}. The output position-position-velocity (PPV) cubes have a spatial resolution of $0.06 \pc$, and a velocity resolution of $0.03 \kms$ {for lines without hyperfine structure (lines with hyperfine structure are discussed below).}

We use collisional rate data taken from the LAMDA \citep{schoier2005} {and EMAA databases}, for the molecules and collision partners listed in Table \ref{tab:moldata}. Where collisional data for both ortho- and para-H$_2$ are available, we assume an ortho:para ratio of $3:1$, and we assume para-NH$_3$ makes up half the total abundance of NH$_3$. As our reaction network does not include separate isotopic chemistry, we assume, where necessary, that the ratio of molecular isotopologues is equal to the underlying isotope ratio, namely $^{12}$C/$^{13}$C$= 77$ and $^{12}$O/$^{18}$O$= 550$ \citep{wilson1994}. Deviations from these ratios under the conditions we investigate are unlikely to be larger than a factor of a few \citep{szucs2014}. We assume thick ice mantle dust opacities from \citet{ossenkopf1994} for wavelengths $\lambda > 1 \um$, and \citet{mathis1983} opacities at shorter wavelengths (see \citealt{clark2012b}).

We produce PPV line intensity cubes for the $J=1-0$ rotational transitions of $^{12}$CO, $^{13}$CO, C$^{18}$O, HCN, N$_2$H$^+$, HCO$^+$ and HNC, and the $J=2-1$ transition of CS. These lines all fall within the $3 \, {\rm mm}$ observational window, and are typically among the brightest in star-forming regions, leading to plentiful observational data as a point of comparison \citep{pety2017,kauffmann2017,barnes2020,tafalla2021,tafalla2023,hacar2024}. We also investigate the para-NH$_3$ $(1,1)$ inversion transition, for which similarly rich observational data is available \citep{ragan2011,ragan2012,friesen2017,feher2022}.

{The NH$_3$, HCN and N$_2$H$^+$ transitions involve multiple hyperfine components, and we use larger velocity resolutions of $0.25$, $0.1$ and $0.1 \kms$ respectively to capture all the structure without increasing the computational cost of the radiative transfer. While {\sc radmc3d} does account for the optical depth effects of multiple overlapping hyperfine transitions, this is not included when determining level populations, meaning that the hyperfine levels are not radiatively coupled with each other. The distribution of flux between the individual components may therefore be calculated incorrectly if this coupling is significant, although we expect the integrated intensities over all hyperfine components to be valid.}

\begin{figure*}
  \centering
  \includegraphics[width=\columnwidth]{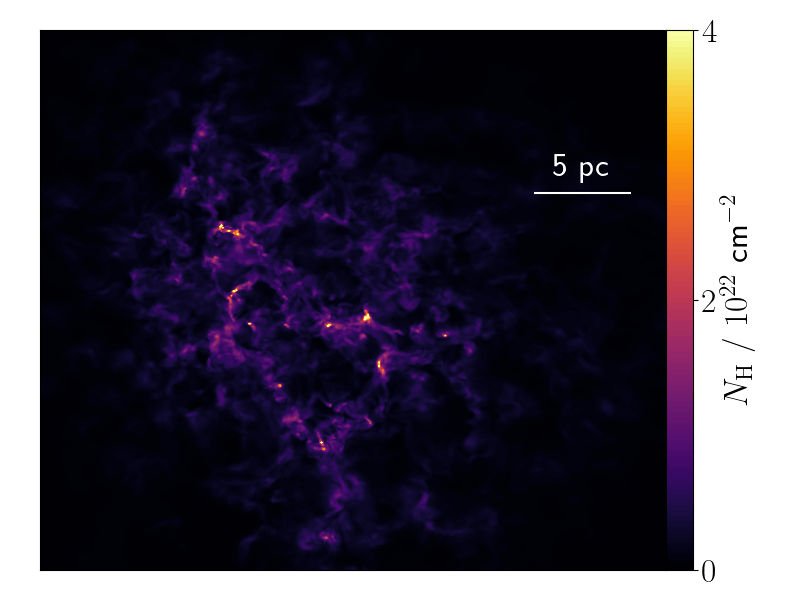}
  \includegraphics[width=\columnwidth]{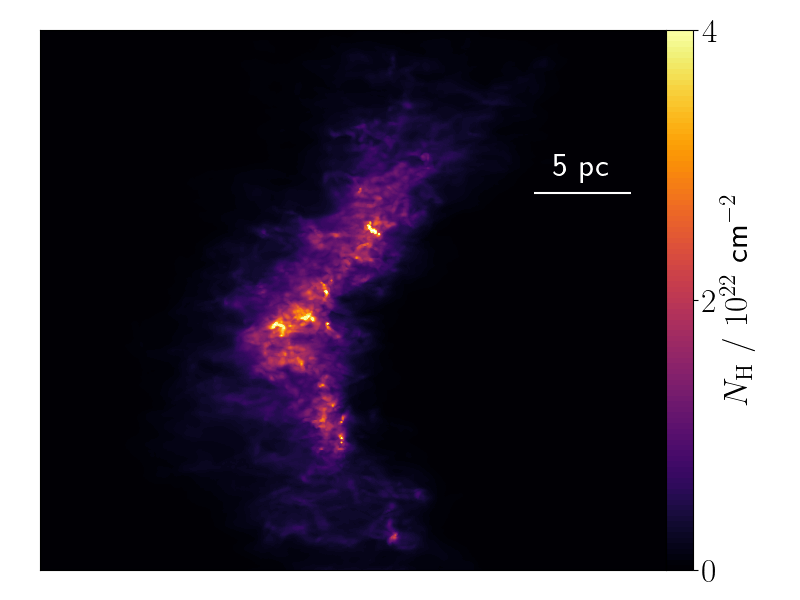}\\
  \caption{Column density maps of the cloud seen face-on (left) and edge-on (right).}
  \label{fig:coldens}
\end{figure*}

\begin{figure*}
  \centering
  \includegraphics[width=0.32\textwidth]{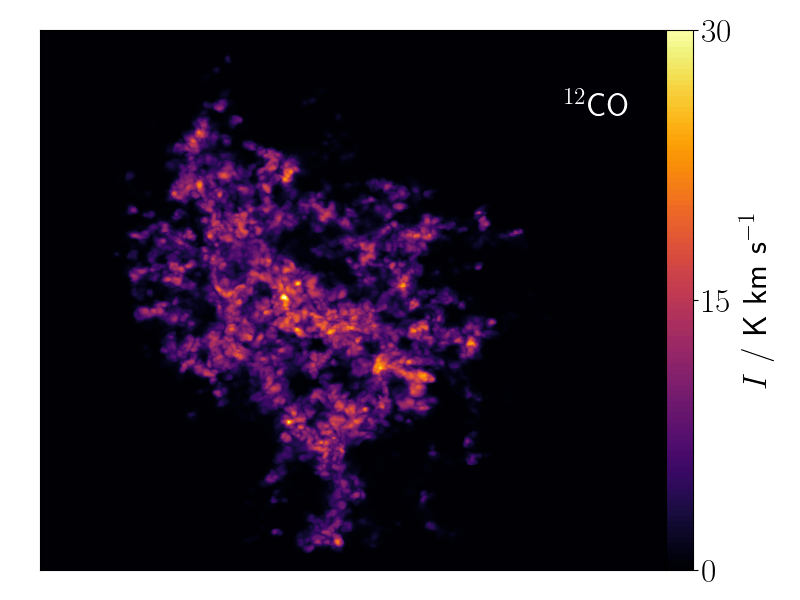}
  \includegraphics[width=0.32\textwidth]{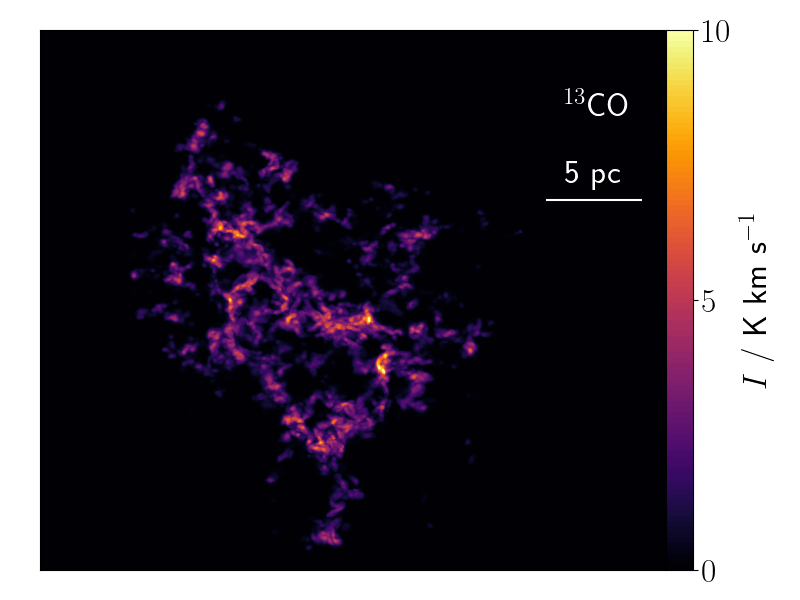}
  \includegraphics[width=0.32\textwidth]{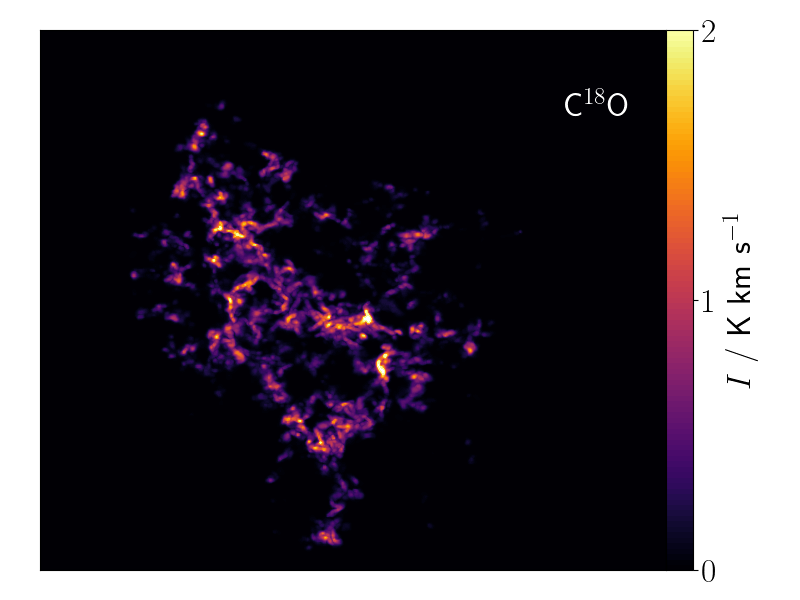}\\
  \includegraphics[width=0.32\textwidth]{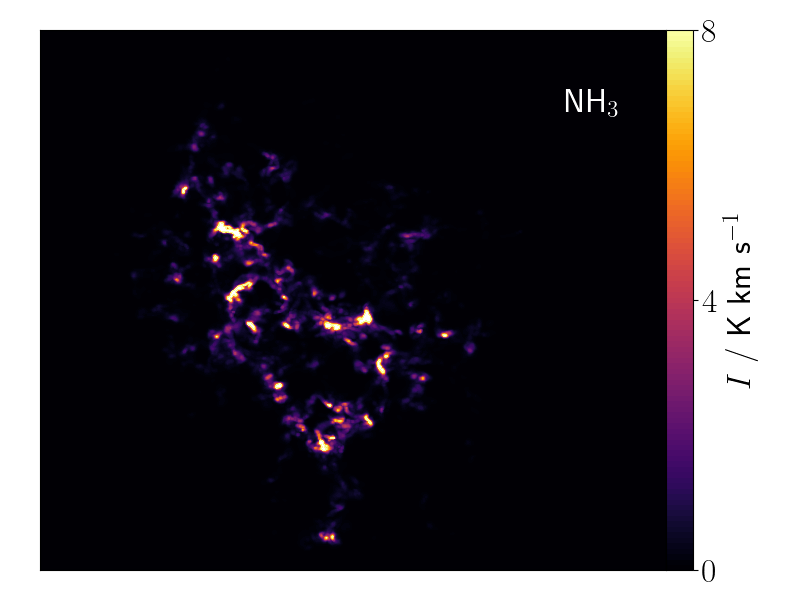}
  \includegraphics[width=0.32\textwidth]{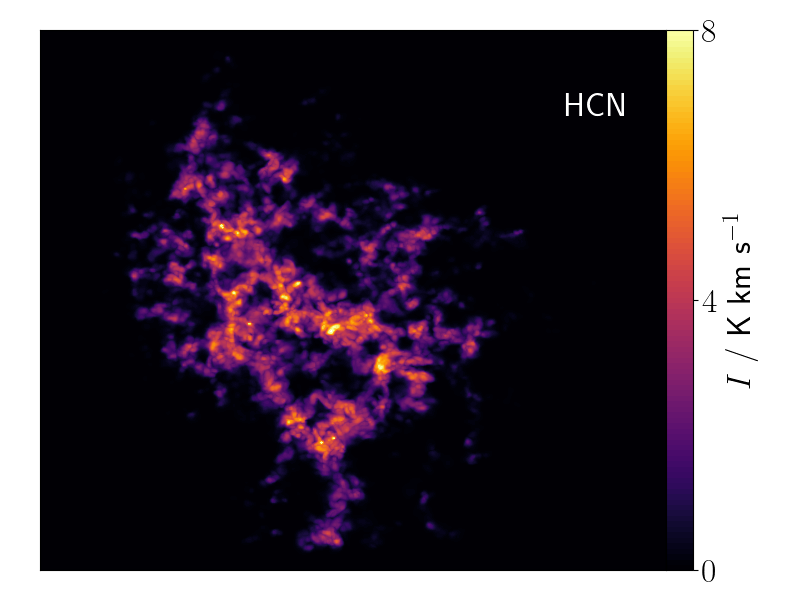}
  \includegraphics[width=0.32\textwidth]{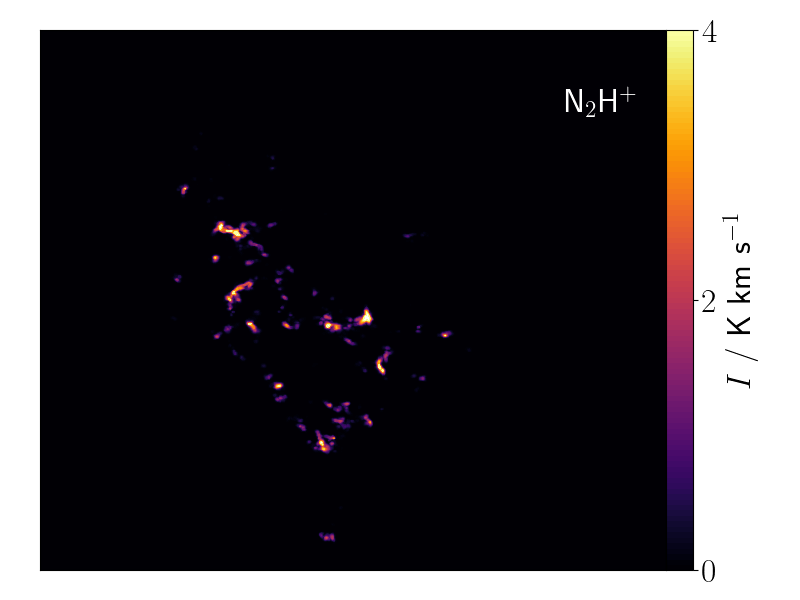}\\
  \includegraphics[width=0.32\textwidth]{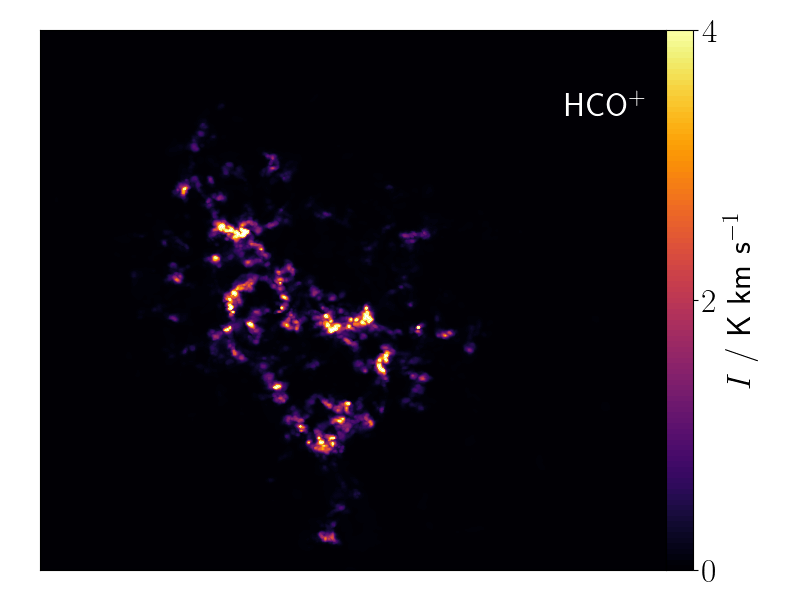}
  \includegraphics[width=0.32\textwidth]{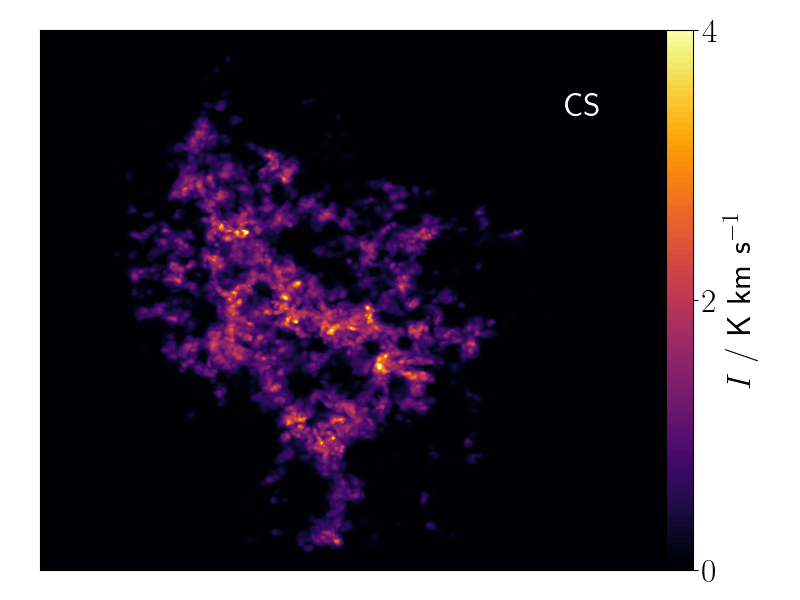}
  \includegraphics[width=0.32\textwidth]{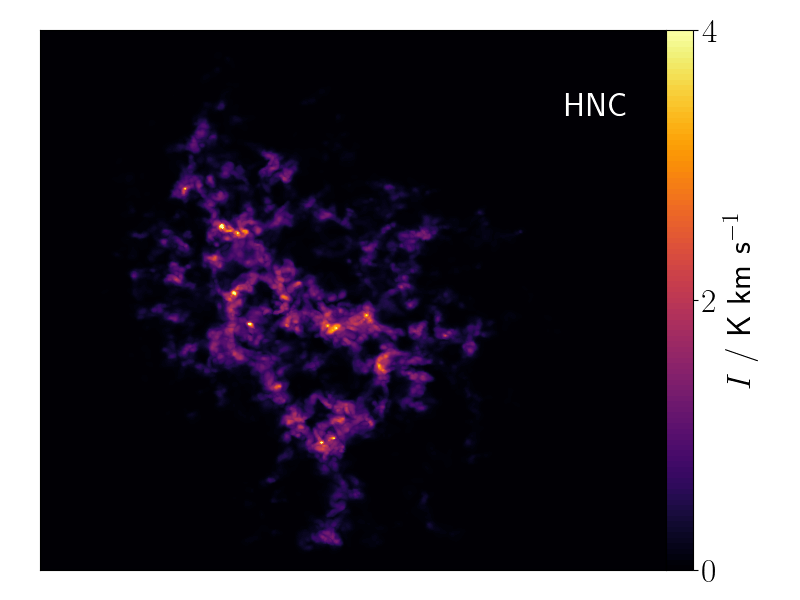}\\
  \caption{Integrated line intensity maps for the cloud seen face-on.}
  \label{fig:linemap}
\end{figure*}

\begin{figure*}
  \centering
  \includegraphics[width=0.32\textwidth]{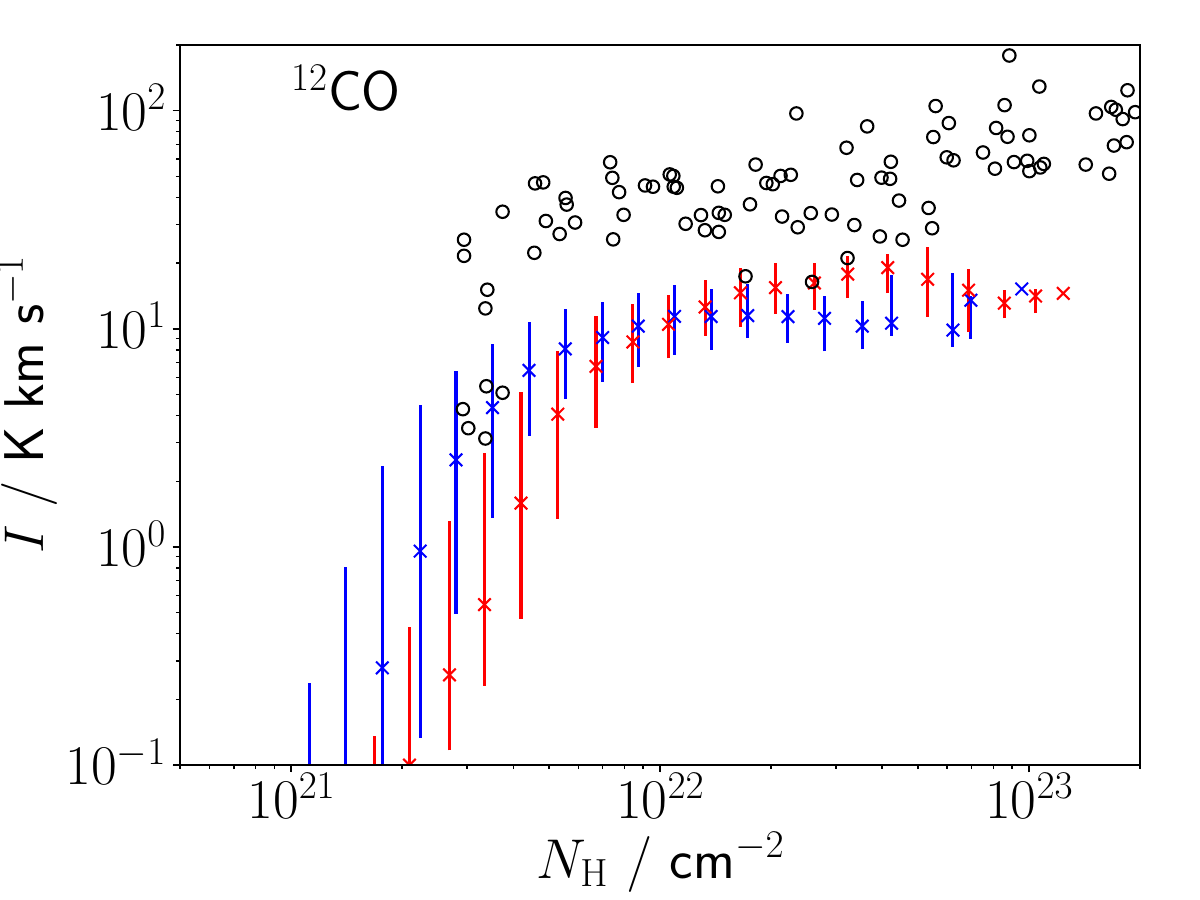}
  \includegraphics[width=0.32\textwidth]{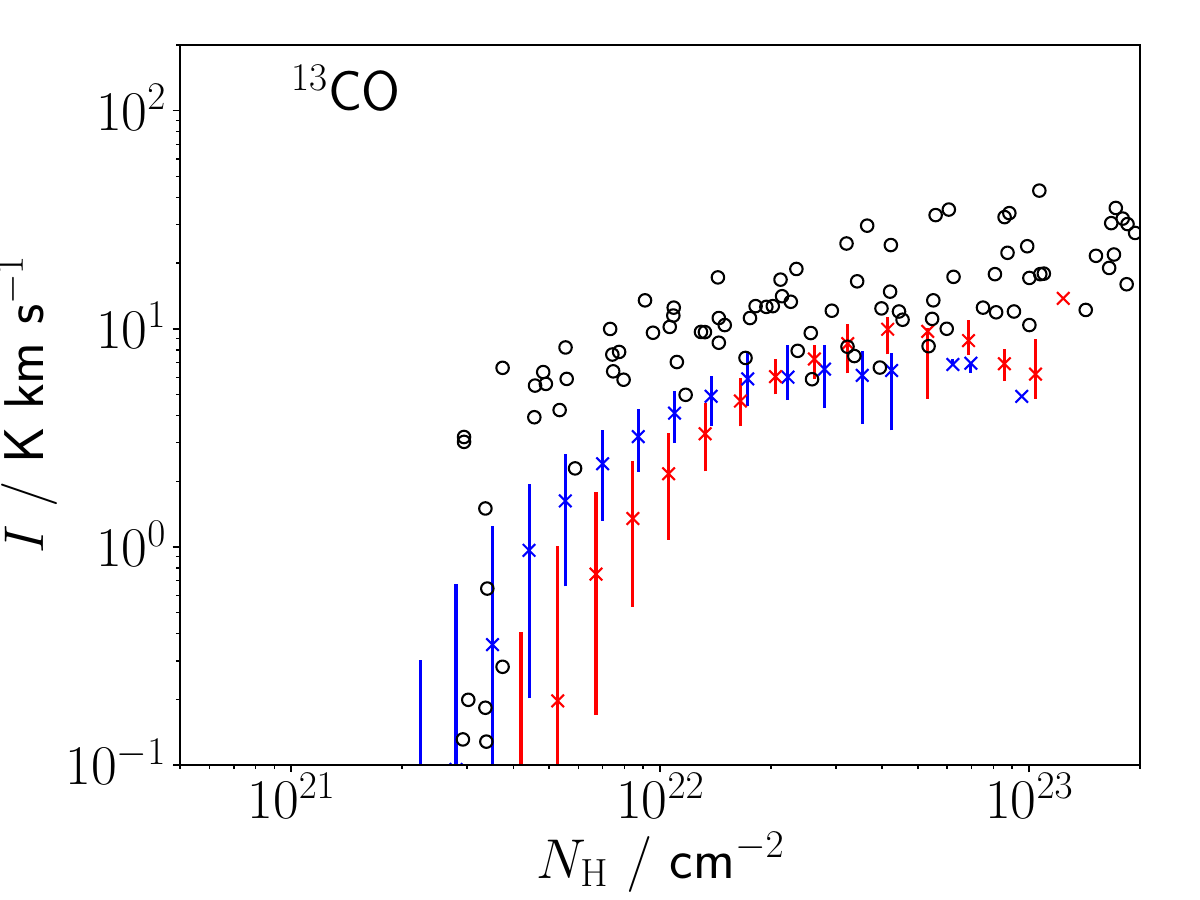}
  \includegraphics[width=0.32\textwidth]{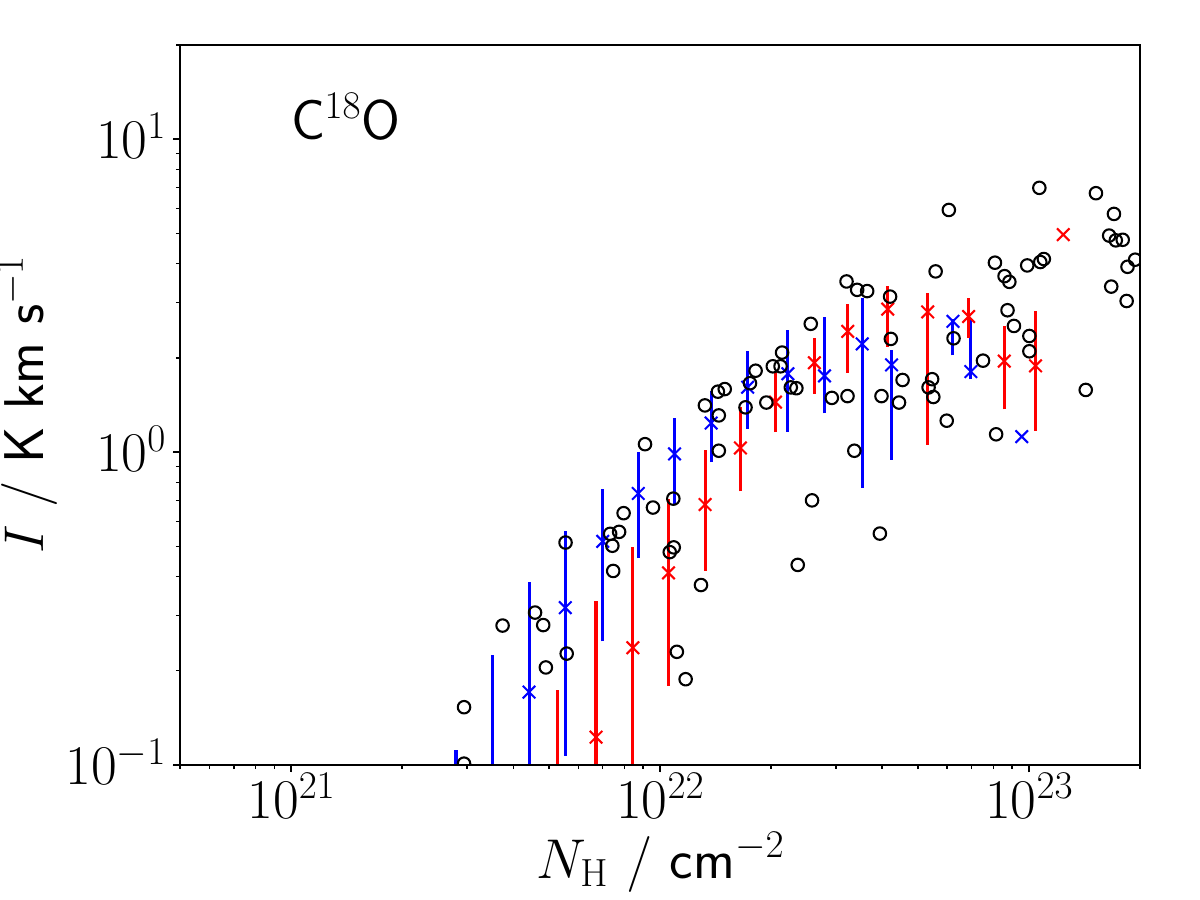}\\
  \includegraphics[width=0.32\textwidth]{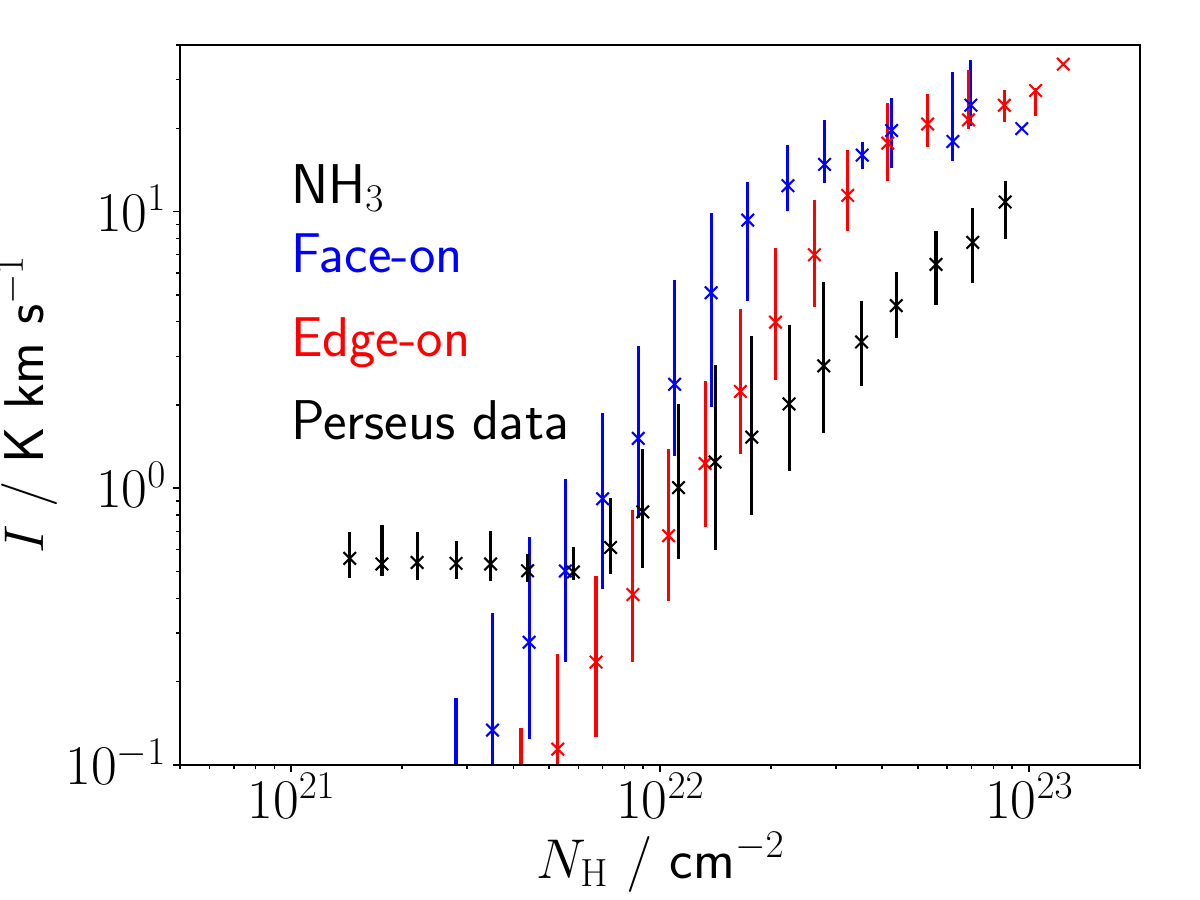}
  \includegraphics[width=0.32\textwidth]{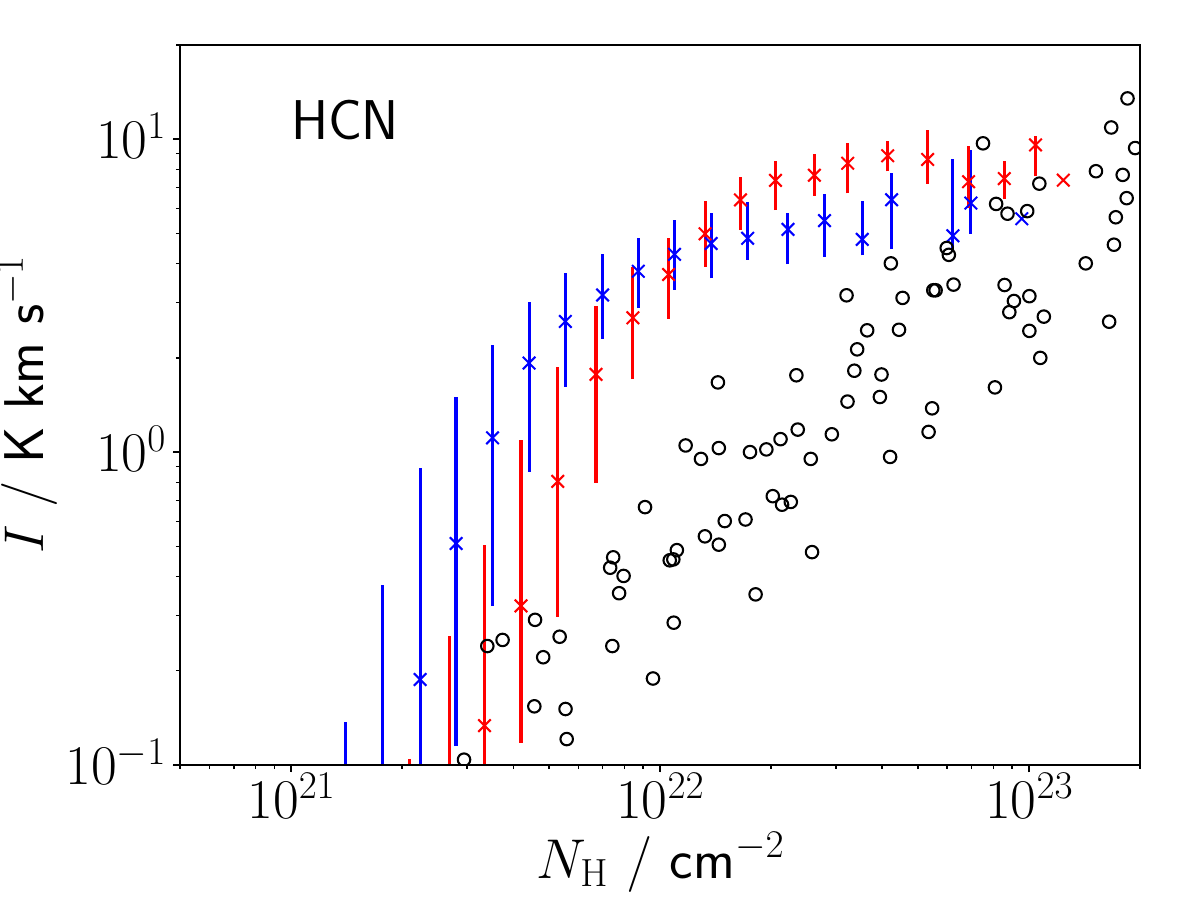}
  \includegraphics[width=0.32\textwidth]{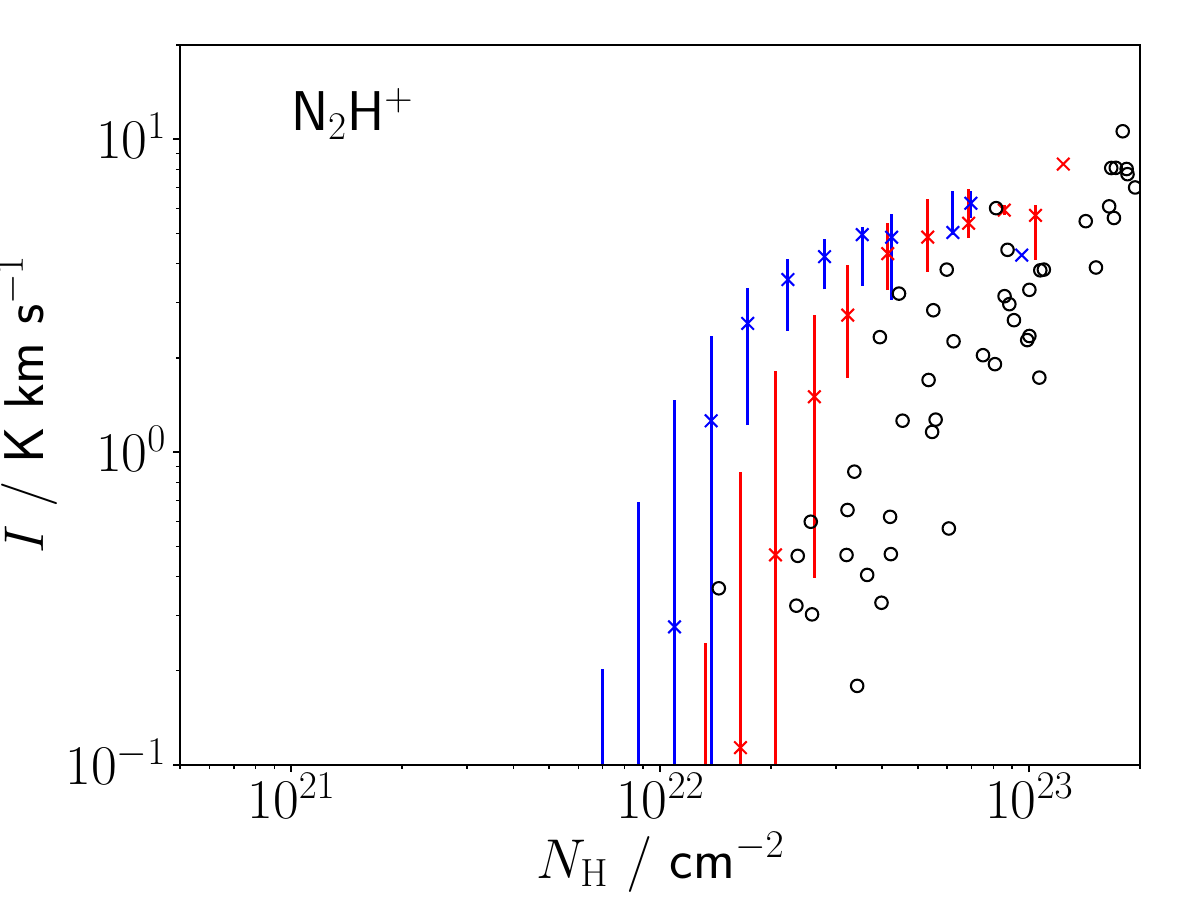}\\
  \includegraphics[width=0.32\textwidth]{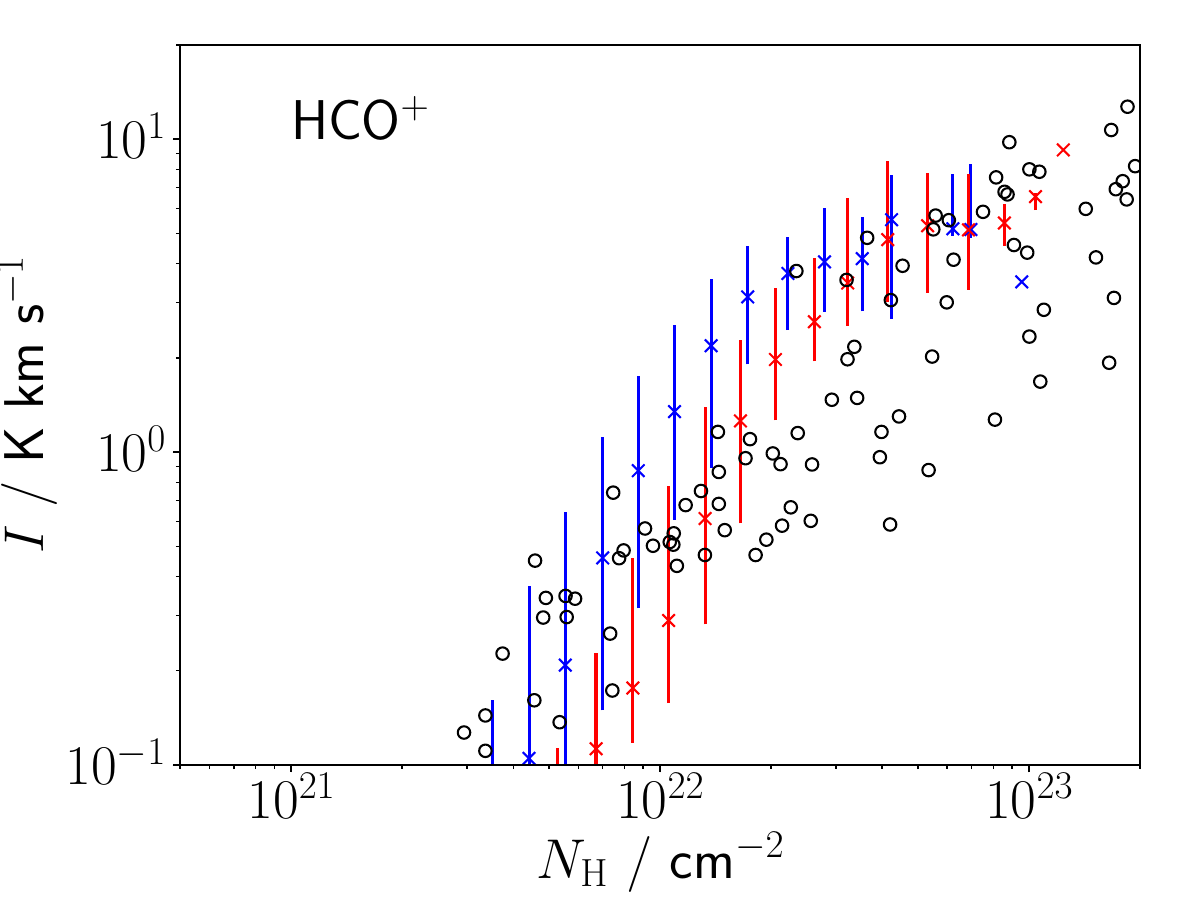}
  \includegraphics[width=0.32\textwidth]{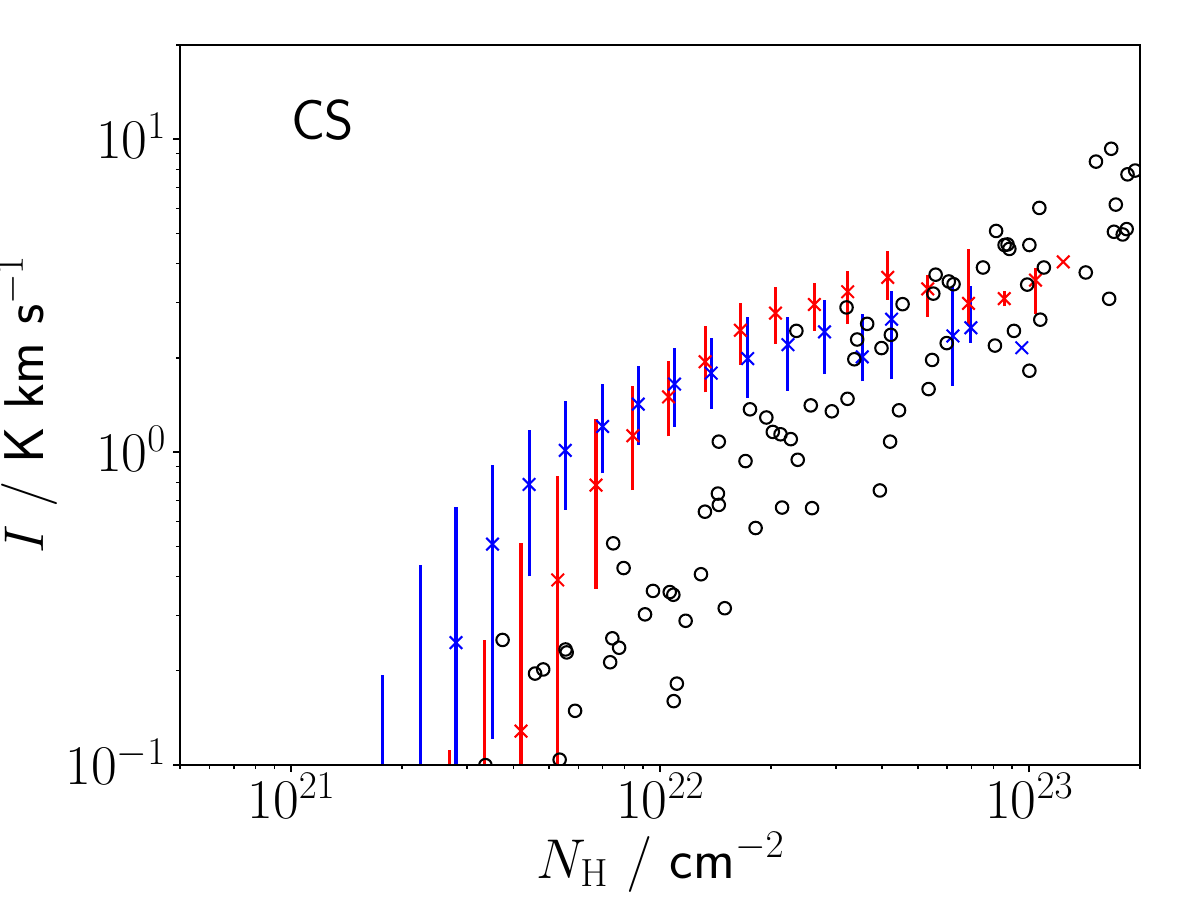}
  \includegraphics[width=0.32\textwidth]{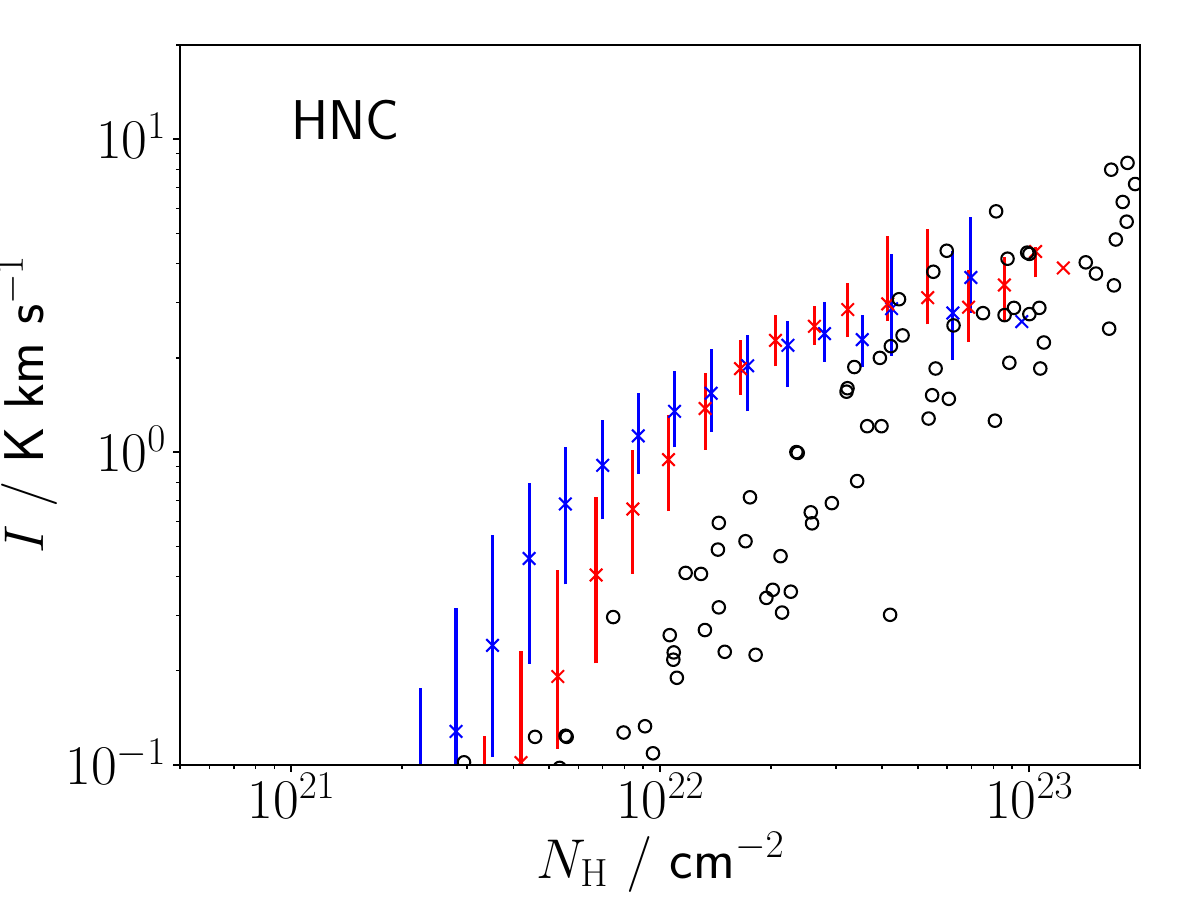}\\
  \caption{Integrated line intensity versus column density for the cloud seen face-on (blue) and edge-on (red). Crosses show the median pixel values, error bars the 16th/84th percentiles. Observational data from the Perseus molecular cloud \citep{friesen2017,tafalla2021} are shown in black.}
  \label{fig:linenh}
\end{figure*}

\begin{figure*}
  \centering
  \includegraphics[width=0.32\textwidth]{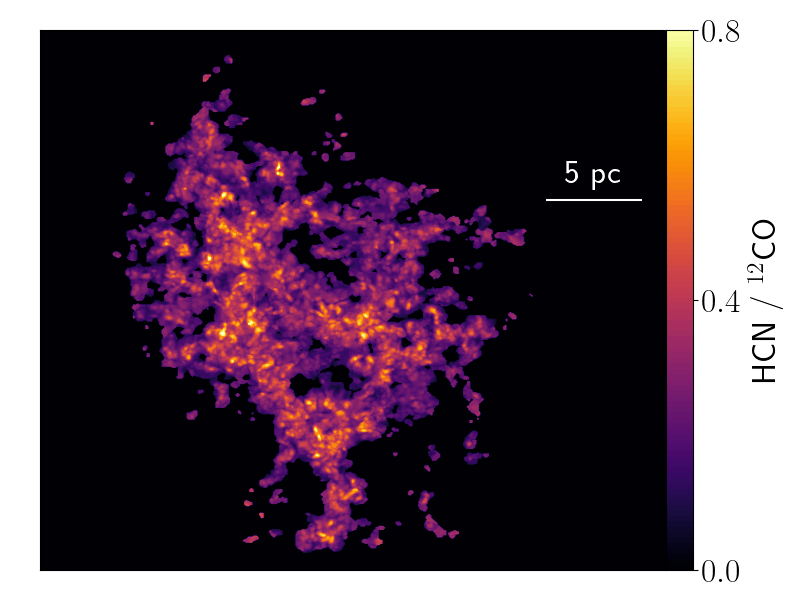}
  \includegraphics[width=0.32\textwidth]{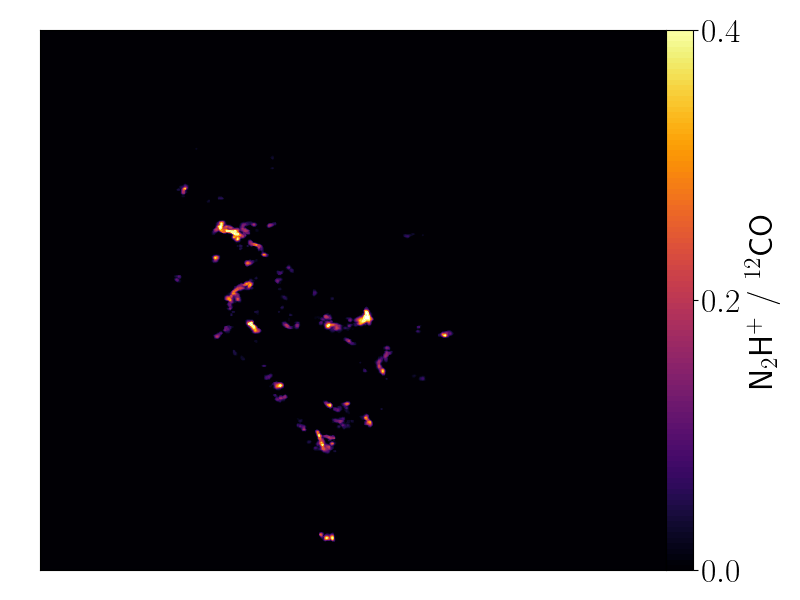}
  \includegraphics[width=0.32\textwidth]{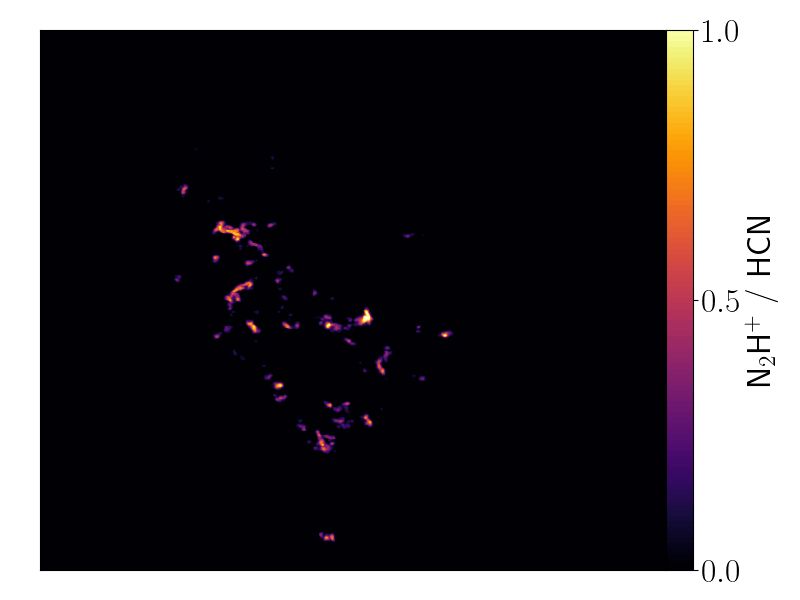}
  \caption{Maps of the ratio of integrated line intensities for HCN and $^{12}$CO (left), N$_2$H$^+$ and $^{12}$CO (centre), and N$_2$H$^+$ and HCN (right), for the cloud seen face-on. Pixels with integrated intensities of either line below $0.2 \kel \kms$ have been masked.}
  \label{fig:lineratios_img}
\end{figure*}

\begin{figure*}
  \includegraphics[width=0.32\textwidth]{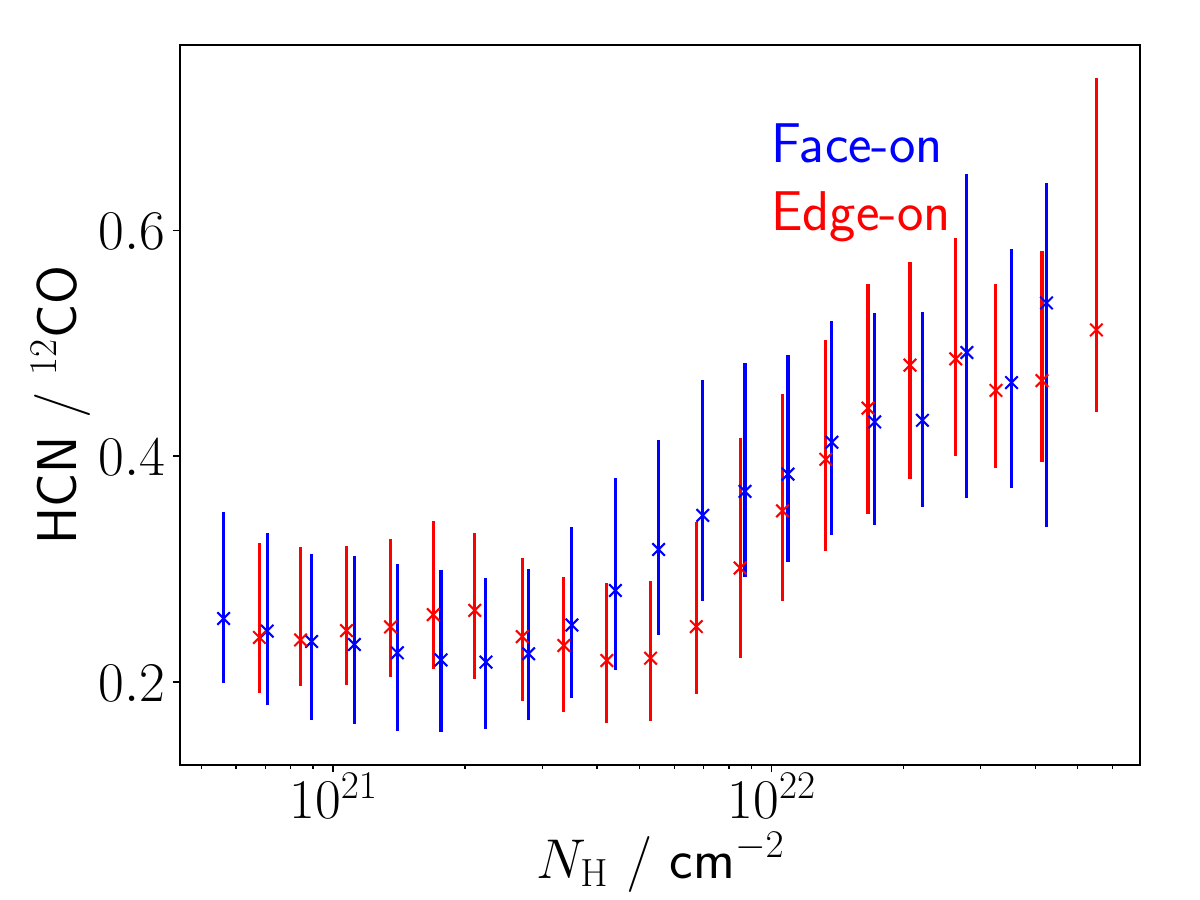}
  \includegraphics[width=0.32\textwidth]{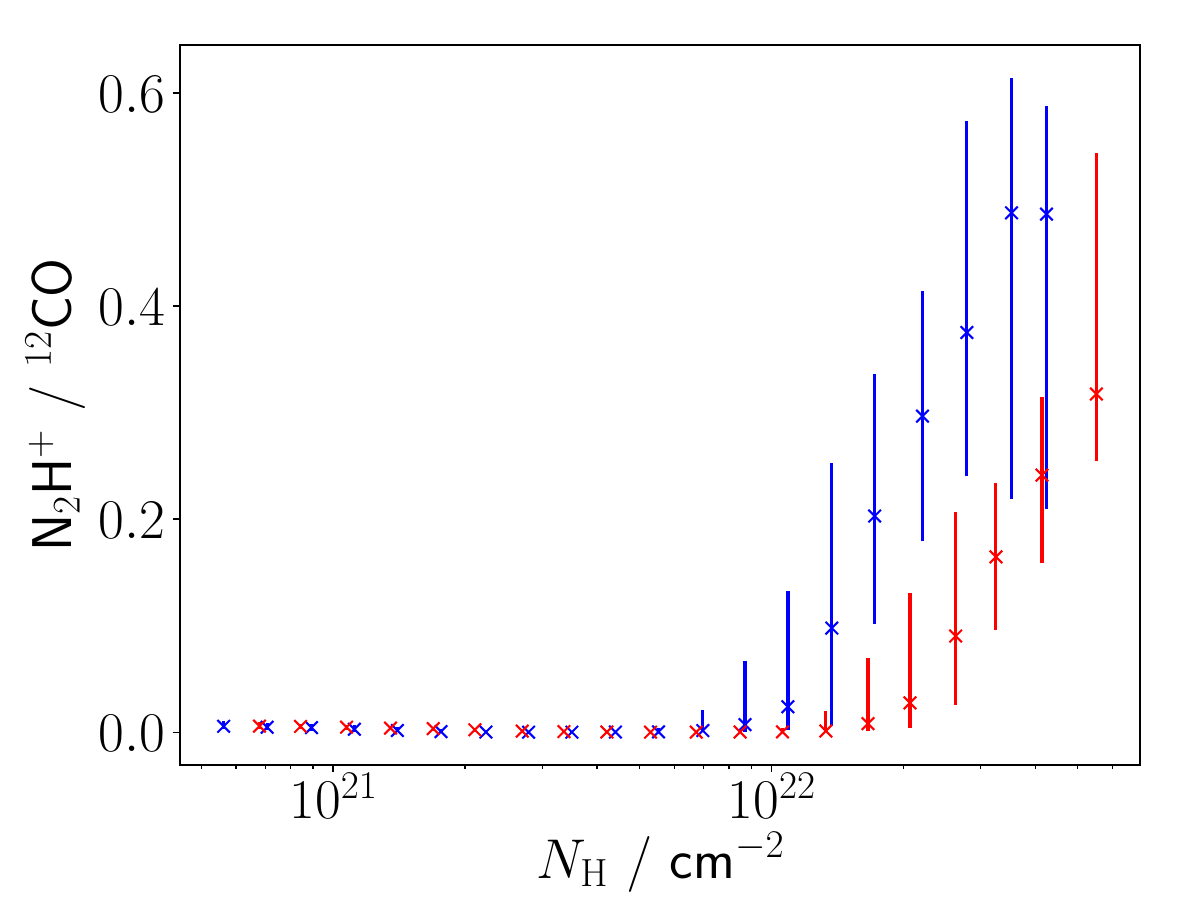}
  \includegraphics[width=0.32\textwidth]{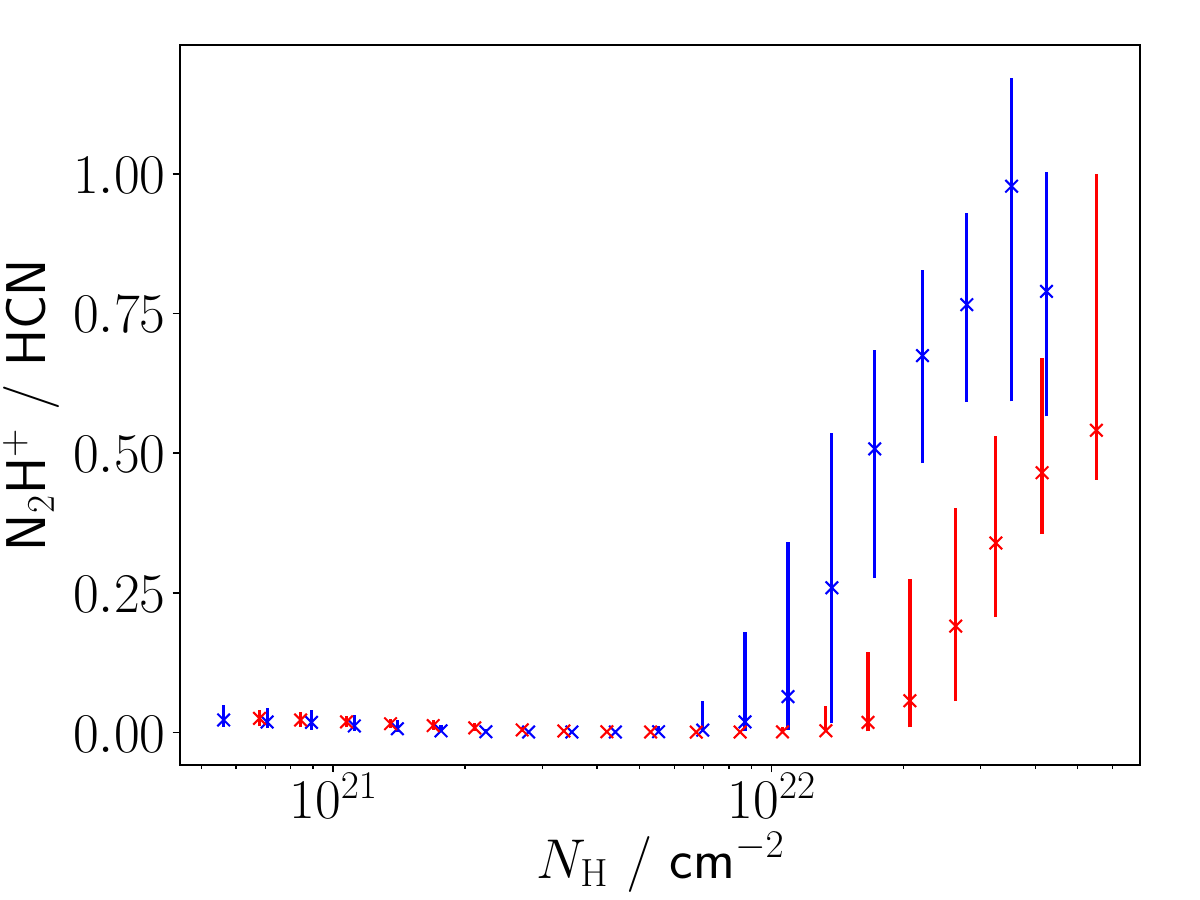}
  \caption{Ratio of integrated line intensities versus column density for HCN and $^{12}$CO (left), N$_2$H$^+$ and $^{12}$CO (centre), and N$_2$H$^+$ and HCN (right), for the cloud seen face-on (blue) and edge-on (red). Crosses show the median pixel values, error bars the 16th/84th percentiles.}
  \label{fig:lineratios_nh}
\end{figure*}

\begin{figure*}
  \centering
  \includegraphics[width=\columnwidth]{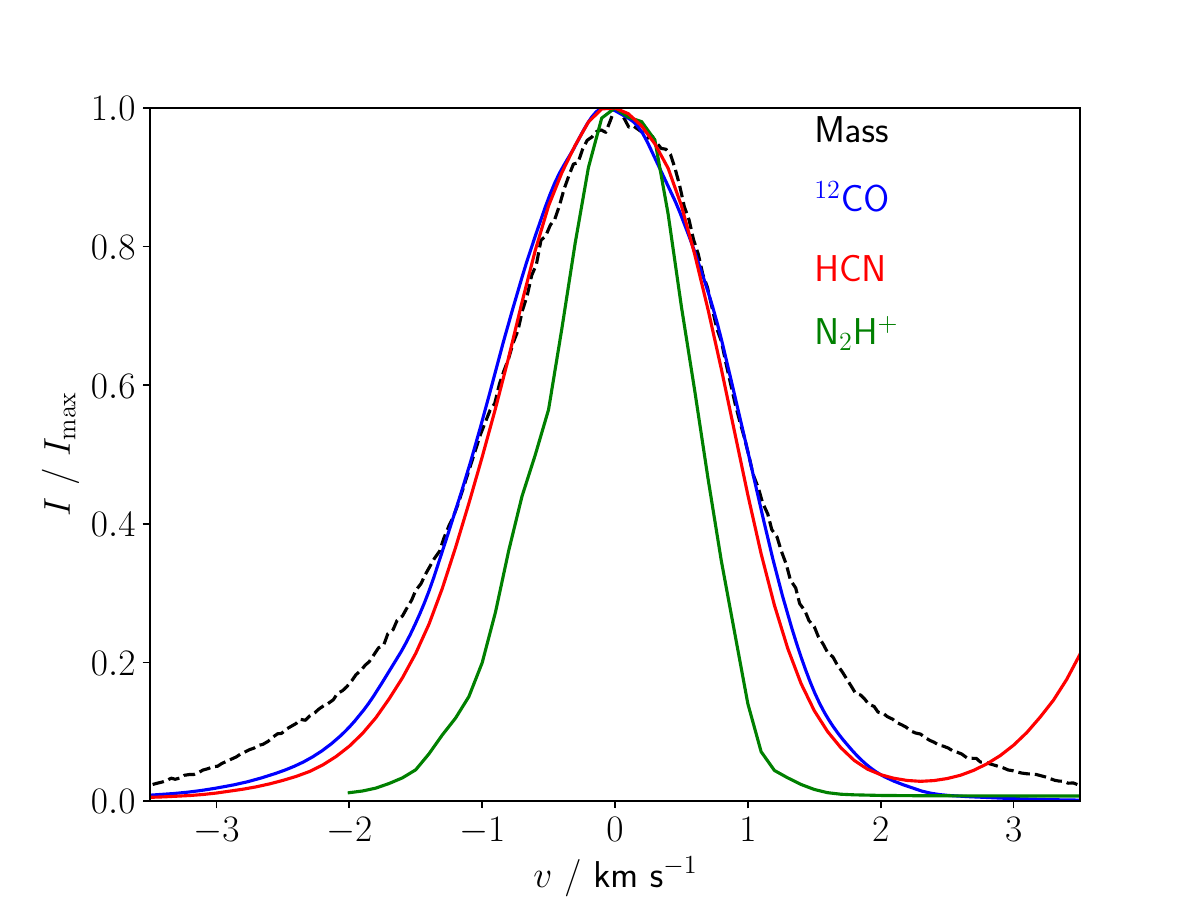}
  \includegraphics[width=\columnwidth]{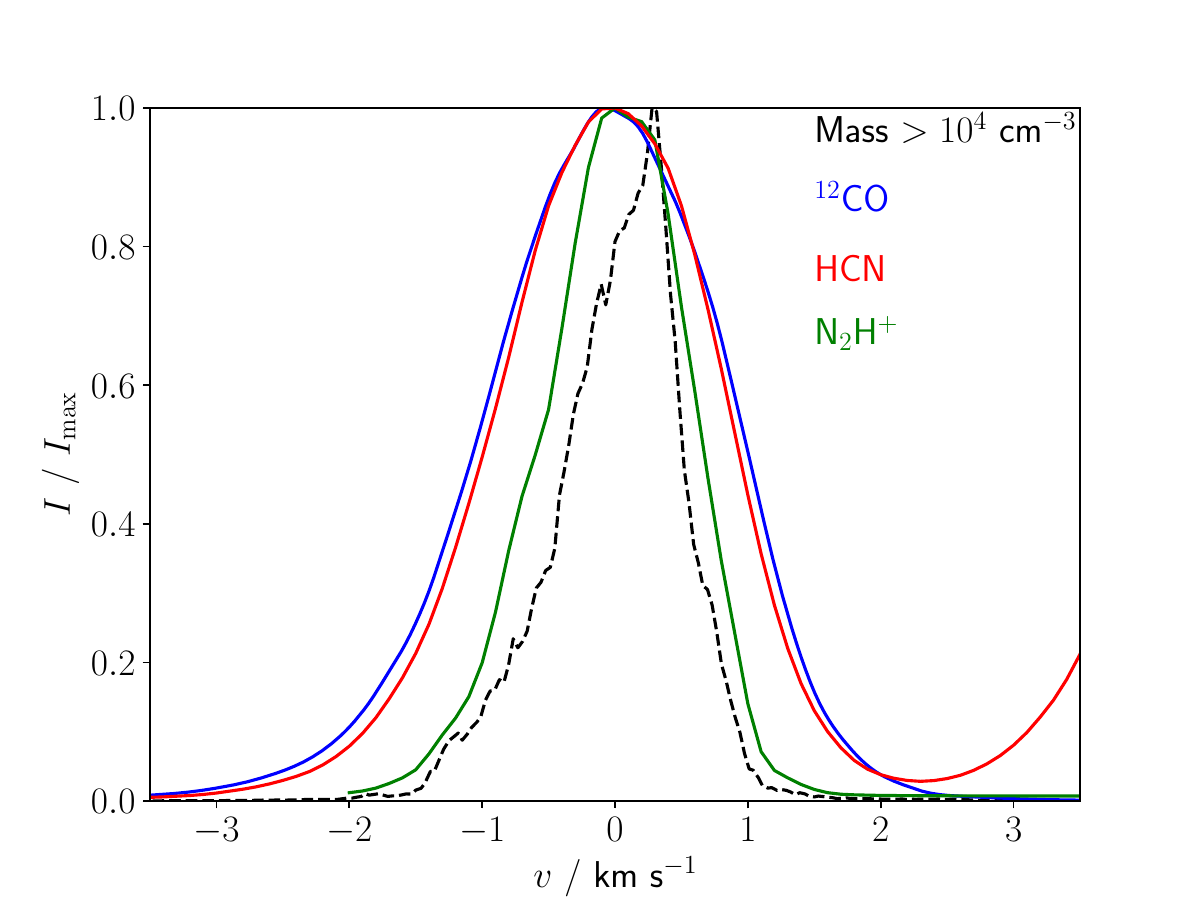}
  \caption{Cloud-averaged line profiles for $^{12}$CO (blue), HCN (red) and N$_2$H$^+$ (green) seen face-on, compared to the velocity distribution of all cloud mass (black, left) and mass above a density of $10^4 \pcc$ (black, right). All quantities have been normalised by their peak values.}
  \label{fig:cloudprofile}
\end{figure*}

\begin{figure}
  \centering
  \includegraphics[width=\columnwidth]{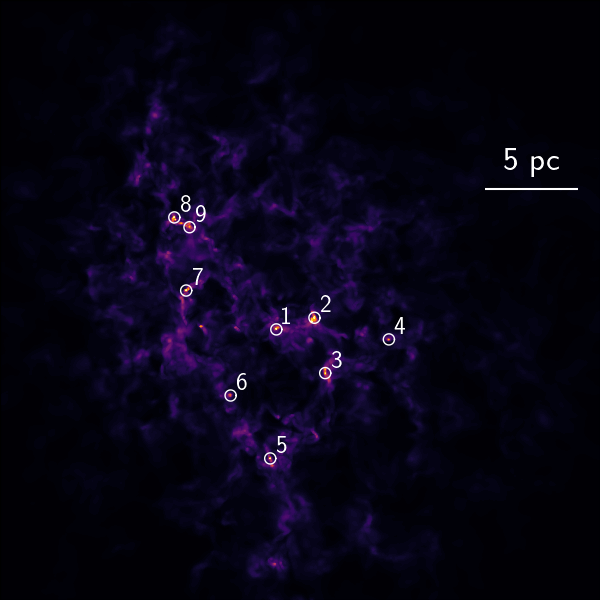}
  \caption{Positions of cores selected from the face-on column density map, highlighted with white circles. Numbers correspond to the labelling in Figure \ref{fig:corelines}. For the purpose of clarity, the circles are three times larger than the $0.1 \pc$ extraction regions.}
  \label{fig:coreimg}
\end{figure}

\begin{figure*}
  \centering
  \includegraphics[width=0.32\textwidth]{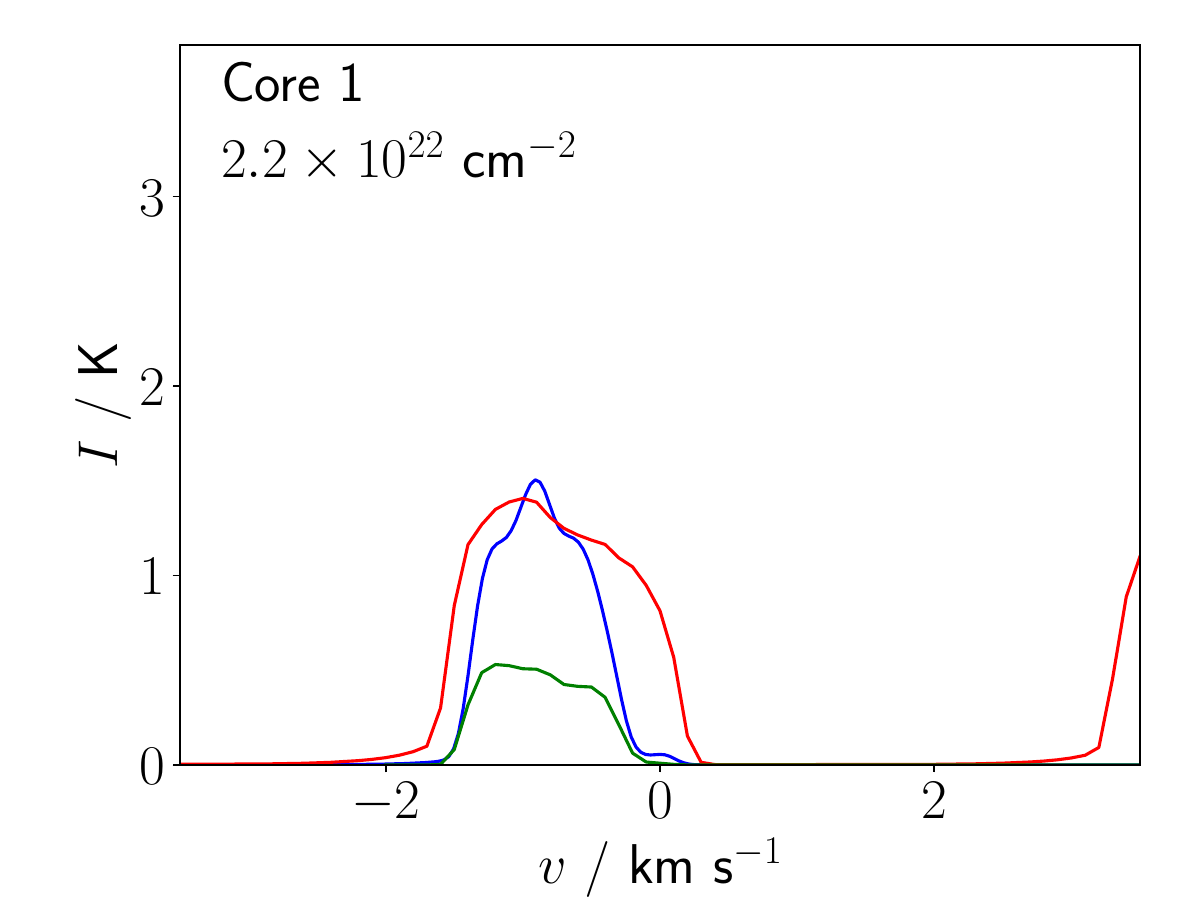}
  \includegraphics[width=0.32\textwidth]{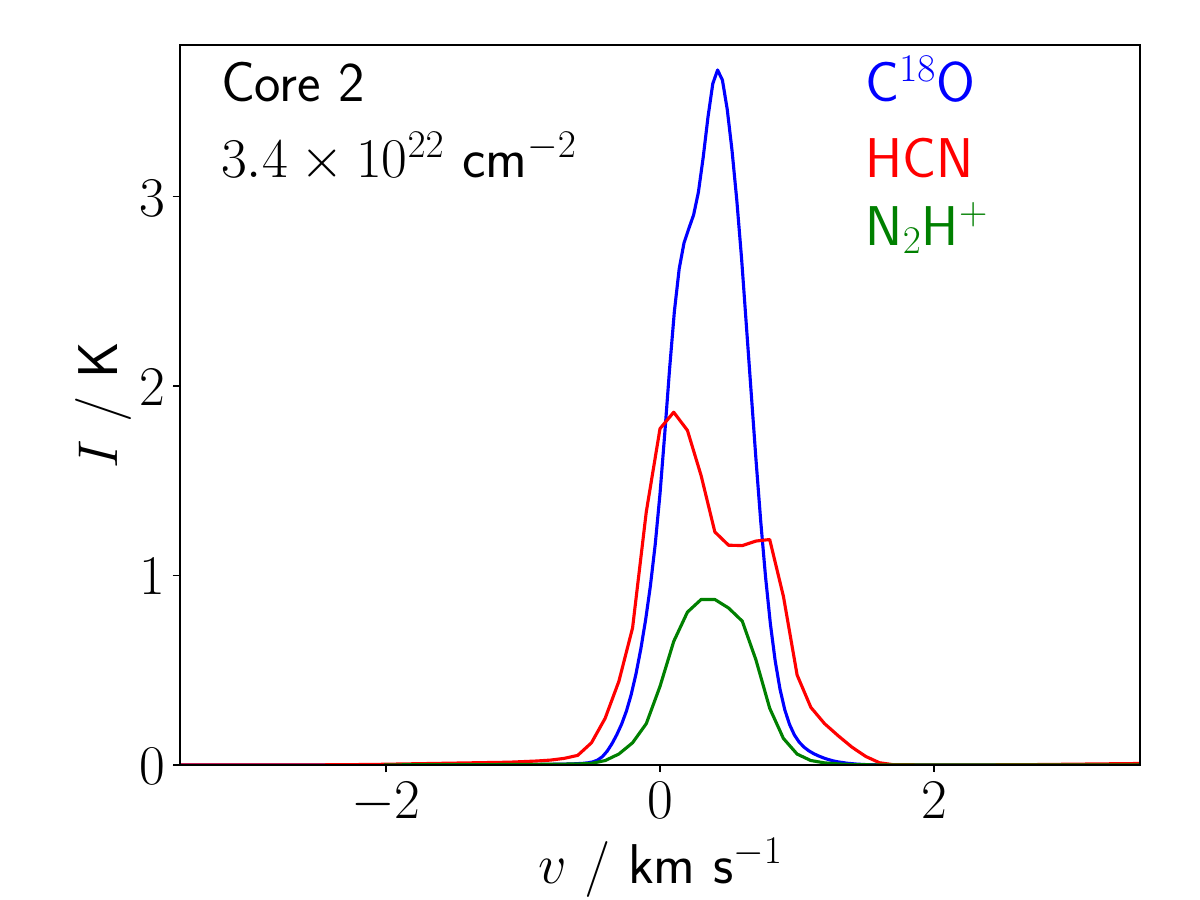}
  \includegraphics[width=0.32\textwidth]{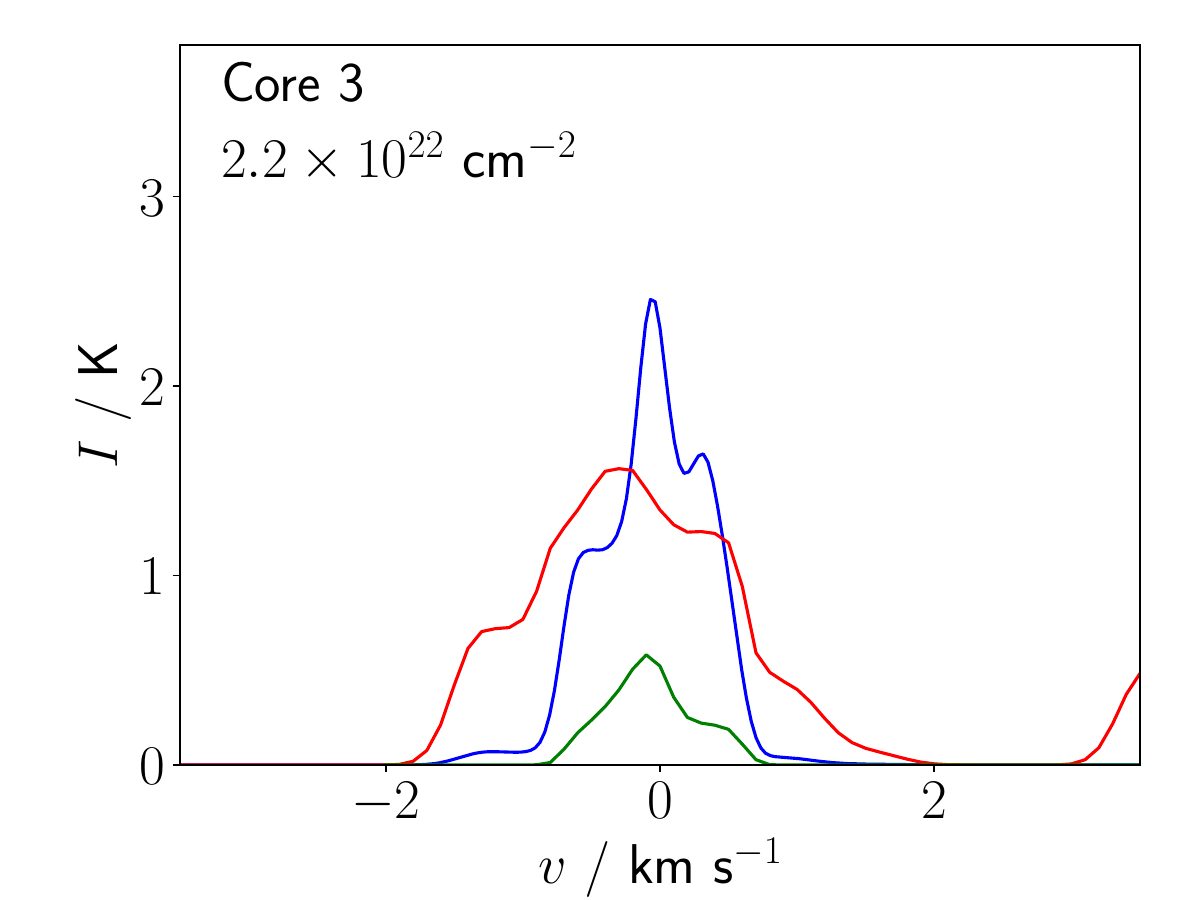}\\
  \includegraphics[width=0.32\textwidth]{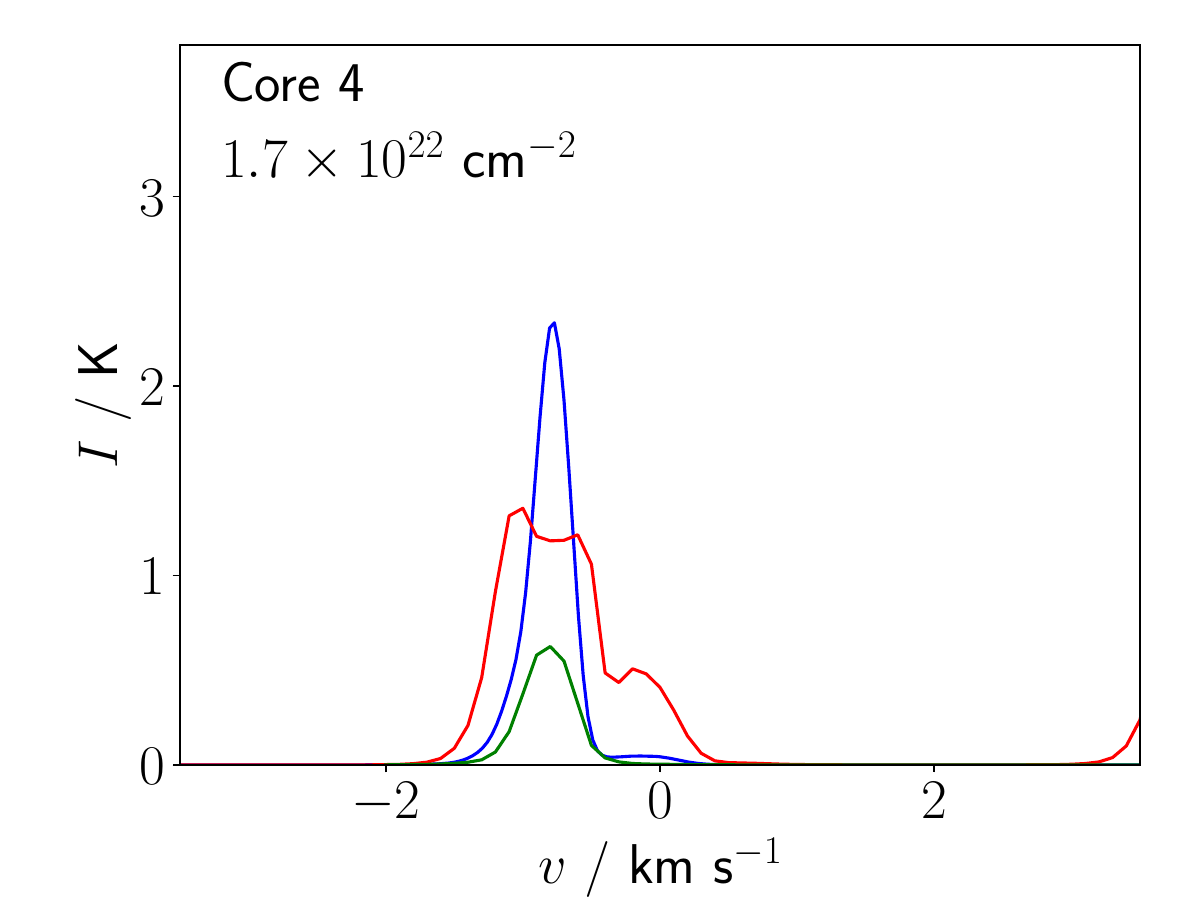}
  \includegraphics[width=0.32\textwidth]{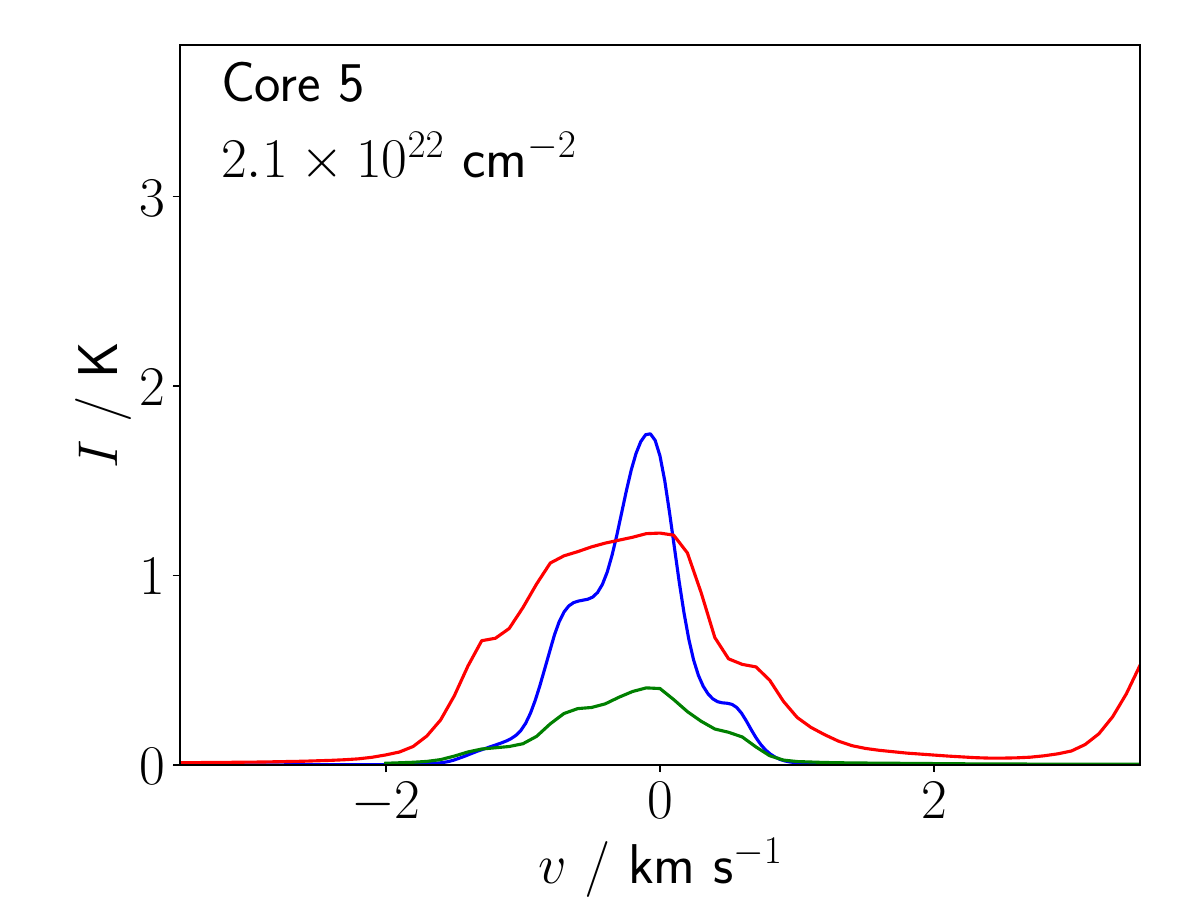}
  \includegraphics[width=0.32\textwidth]{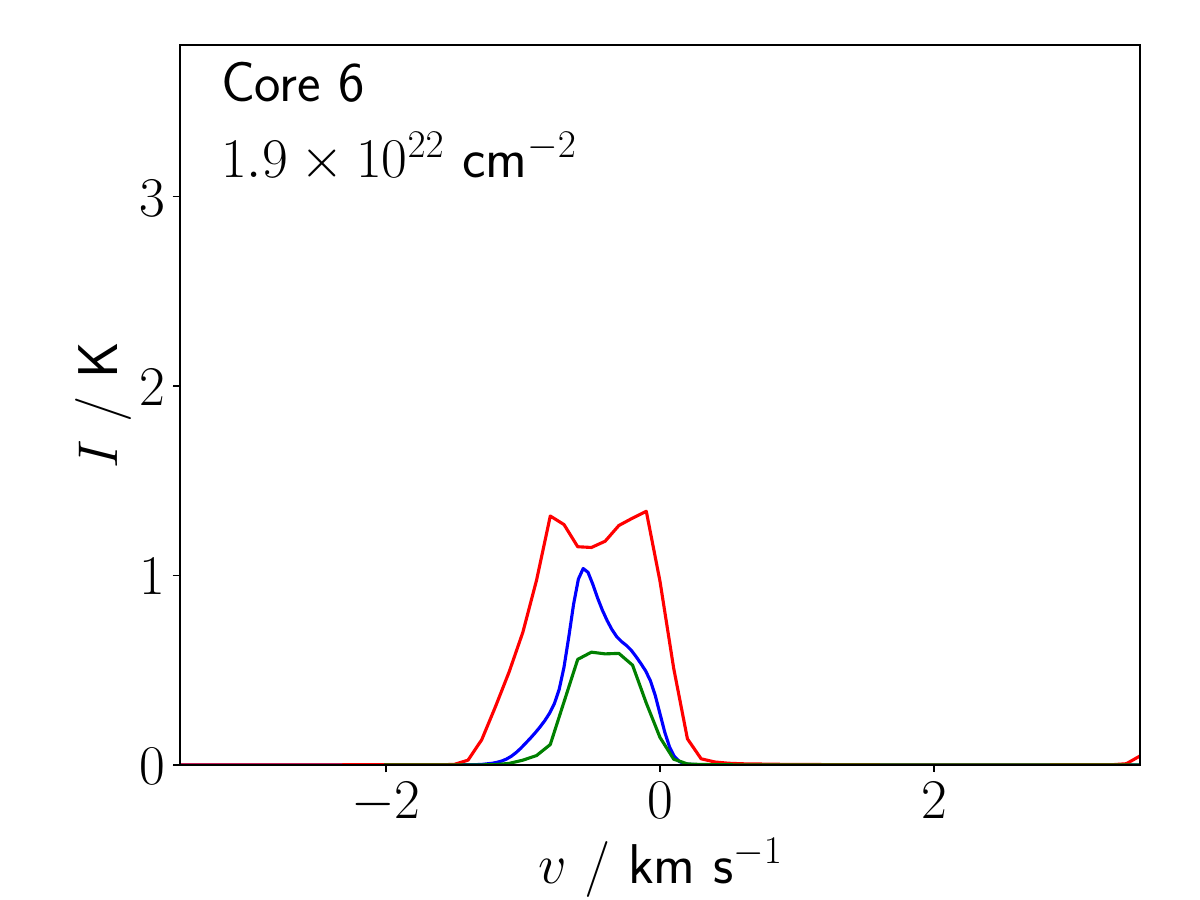}\\
  \includegraphics[width=0.32\textwidth]{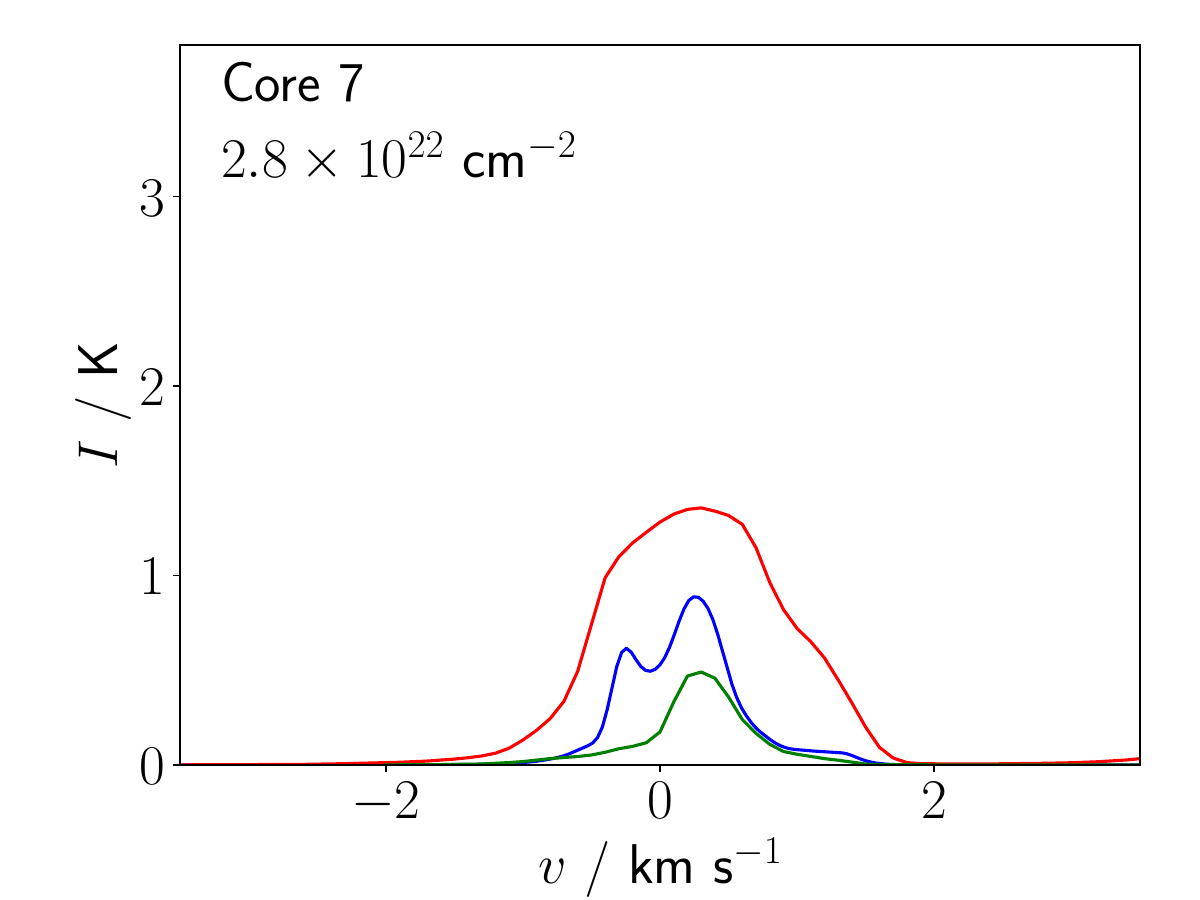}
  \includegraphics[width=0.32\textwidth]{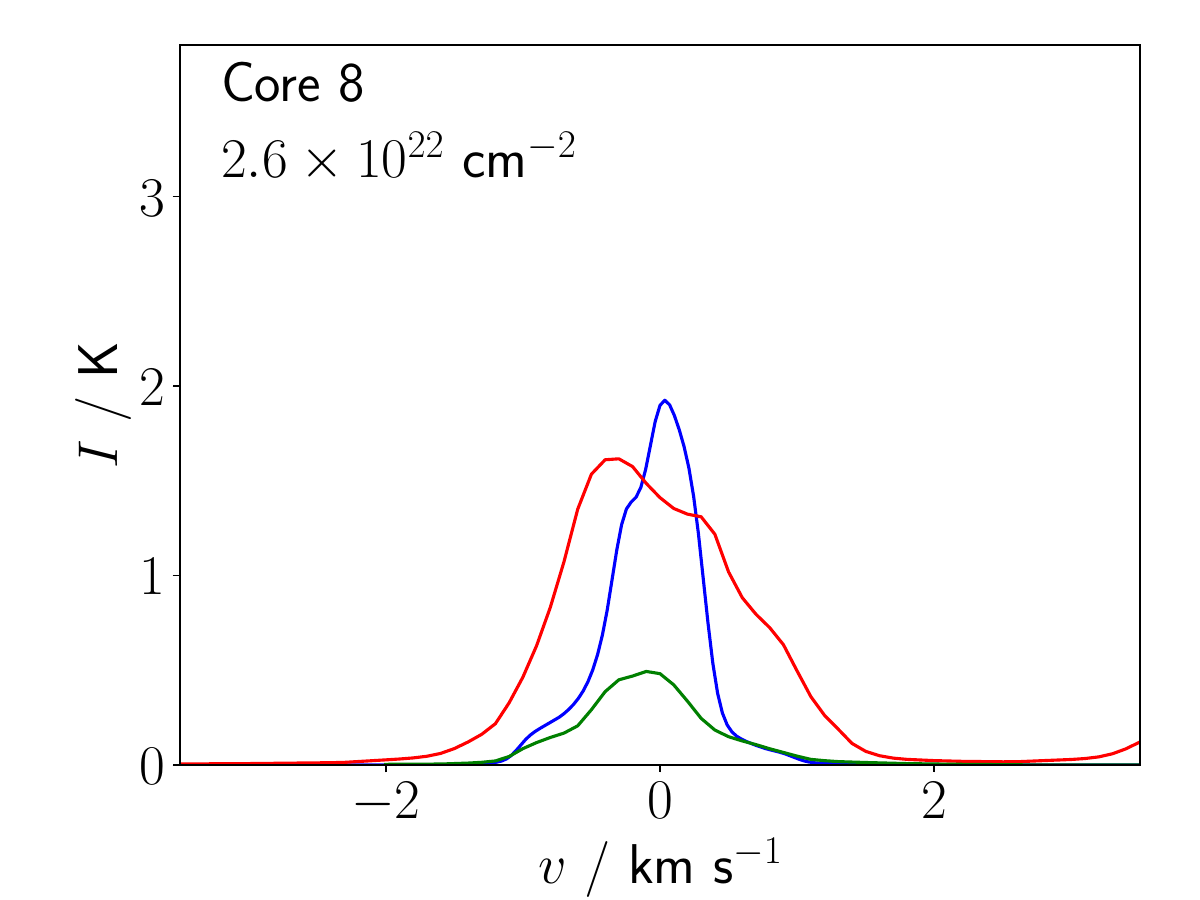}
  \includegraphics[width=0.32\textwidth]{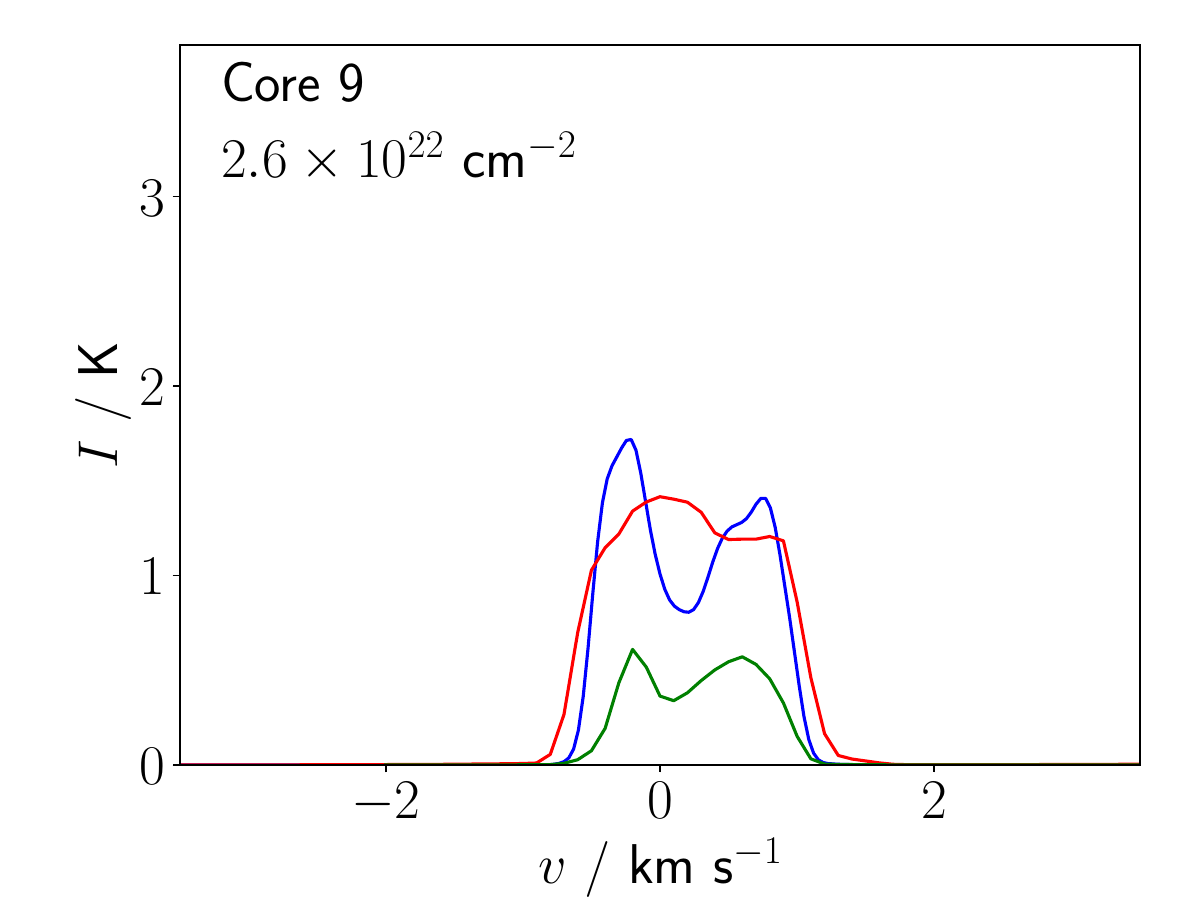}\\
  \caption{Line profiles of C$^{18}$O (blue), HCN (red) and N$_2$H$^+$ (green) for the cores identified from the simulated cloud in Figure \ref{fig:coreimg}. Text labels indicate the column density of the $0.1 \pc$ extraction region.}
  \label{fig:corelines}
\end{figure*}

\begin{figure*}
  \centering
  \includegraphics[width=\columnwidth]{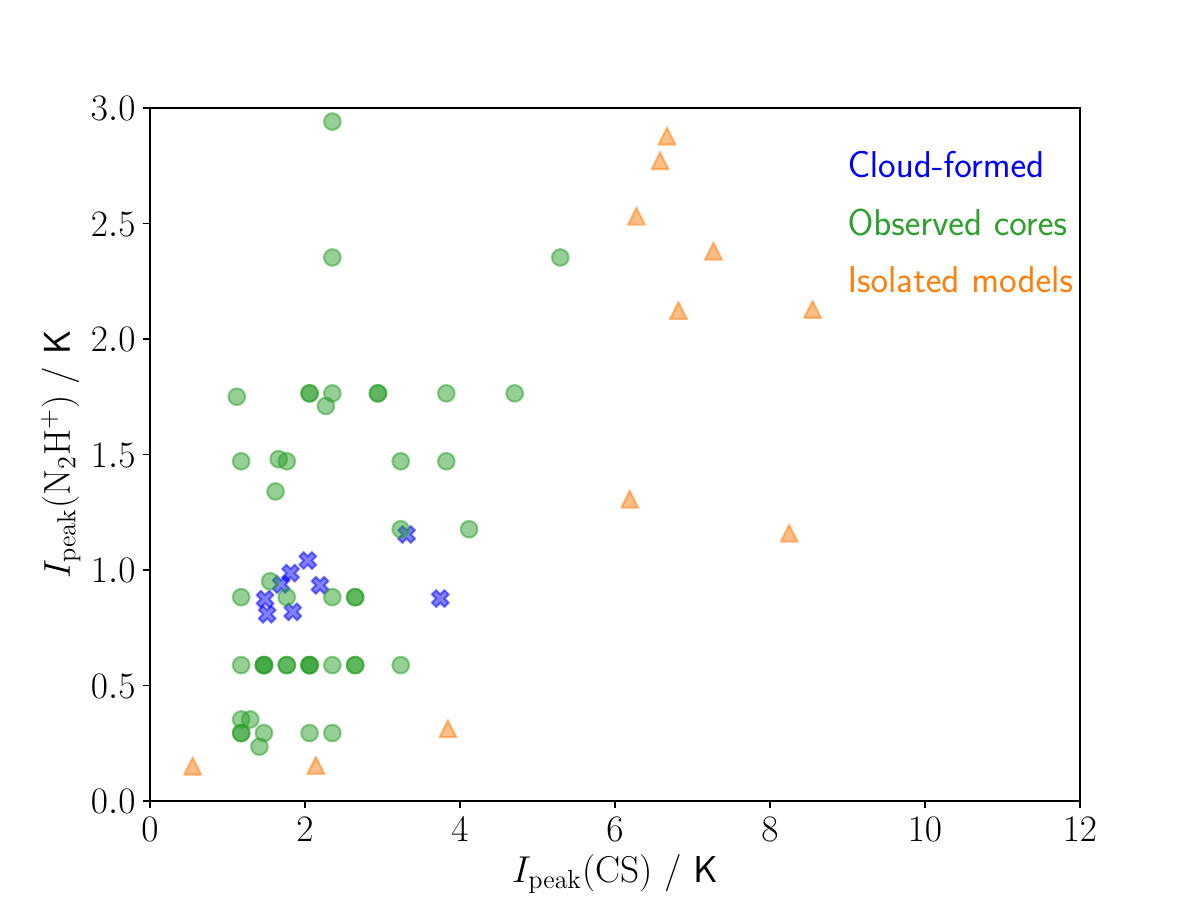}
  \includegraphics[width=\columnwidth]{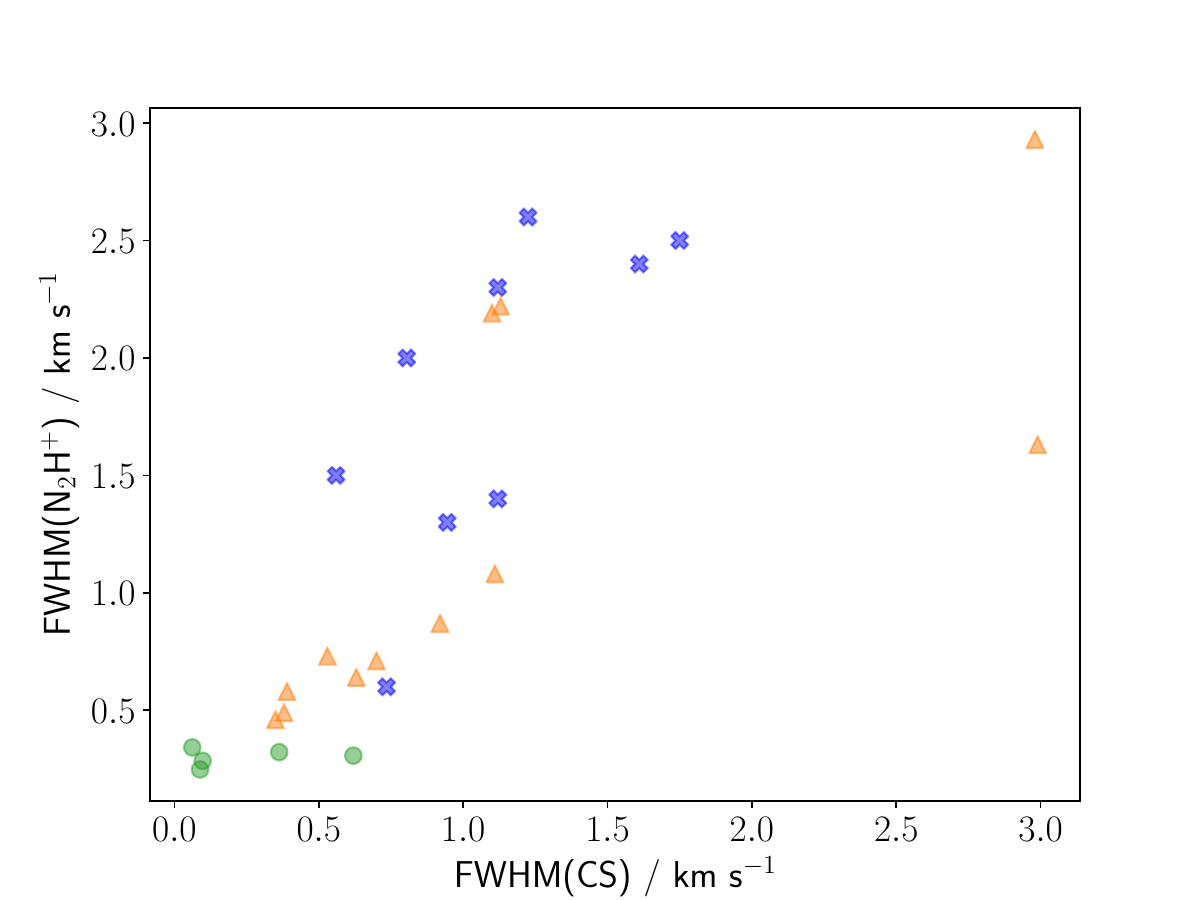}
  \caption{Distribution of the peak line intensities (left) and FWHMs (right) of CS versus N$_2$H$^+$. Blue crosses show the cores identified from the simulated cloud in Figure \ref{fig:coreimg}. Green circles show observational data taken from \citet{lee1999} and \citet{tafalla2002}. Orange triangles show the results of initially-spherical core models from \citet{priestley2022}.}
  \label{fig:p22}
\end{figure*}

\section{Results}

\subsection{Cloud-scale line properties}

Figure \ref{fig:coldens} shows column density maps {in total hydrogen nuclei of} the simulated cloud, viewed both parallel (`face-on') and perpendicular (`edge-on') to the collision axis. The collision between the initial atomic clouds forms a layer of enhanced density at their interface, which subsequently fragments into a network of filaments and cores. The initial turbulent velocity field causes deviations from spherical symmetry, such as the `bend' in the cloud when viewed edge-on. Figure \ref{fig:linemap} shows integrated intensity maps of lines from the molecules listed in Table \ref{tab:moldata}, for the face-on cloud orientation (line emission maps of the edge-on cloud show the same qualitative behaviour).

The CO isotopologues, with their relatively high abundances and low critical excitation densities, show widespread emission throughout the cloud, although this is less pronounced for the rarest (C$^{18}$O). This behaviour is also seen for several species commonly thought of as tracers of high-density material, most notably HCN, CS and HNC, as is found observationally \citep{pety2017,kauffmann2017,evans2020}. The line emission maps of NH$_3$ and HCO$^+$ more closely track the actual column density of material shown in Figure \ref{fig:coldens}, while N$_2$H$^+$ is the only species investigated which selectively traces the densest regions of the cloud. These species' tendency to trace higher density gas than CO is due to chemical differences causing their abundances to peak at higher volume densities, whereas species such as CS and HCN are chemically quite similar to CO \citep{priestley2023c}.

Figure \ref{fig:linenh} shows the relationships between integrated line intensity and column density, compared to observations of the Perseus molecular cloud. For all lines except NH$_3$, the data are the pointed observations taken by \citet{tafalla2021}. For NH$_3$, we use maps of the NGC 1333 subregion from \citet{friesen2017}, which have been re-reduced (Pineda et al. in prep.) and matched to the {\it Herschel}-derived column density\footnote{{Converted from H$_2$ column to total hydrogen column assuming all hydrogen is in the form of H$_2$.}} maps of \citet{singh2022}. The noise level of the NH$_3$ observations is $\sim 0.5 \kel$, so the apparently-constant intensity at around this value below a column density of $10^{22} \pcs$ should be interpreted as an upper limit. We note that the commonly-used approach of fitting single-temperature modified blackbodies to far-IR data may introduce an observational bias in the measured column densities, but in Appendix \ref{sec:dust} we demonstrate that this effect is modest (a factor of a few at the highest column density values), so we use the true column densities from the simulation throughout the rest of this paper.

The predicted relationships between column density and line intensity for the three CO isotopologues are in general agreement with those observed, although the $^{12}$CO line is somewhat weaker than observed. {A possible explanation for this difference is that} our assumed radiation field strength of $1.7$ Habing may be too low to properly represent the Perseus molecular cloud; \citet{tafalla2023} find that regions with more active star formation produce more $^{12}$CO emission at a given column density, which they attribute to higher gas temperatures caused by a stronger local UV field. The predicted strengths of the $^{13}$CO and C$^{18}$O lines, which originate from further within the cloud, are in better agreement with the data, suggesting that this effect, {if relevant}, is only important in the outer, less-shielded regions.

{Alternatively, the difference in $^{12}$CO intensities may be due to systematic differences in the velocity dispersions of our simulated cloud and Perseus. \citet{tafalla2021} find typical $^{12}$CO full-widths at half-maximum (FWHMs) of $3-4 \kms$ in Perseus, with little column density variation. Our simulated cloud also has a flat FWHM-column density relationship, but at a lower value of $\sim 1 \kms$. If the velocity dispersion were to be increased to match the Perseus FWHM values, the increase in integrated intensity could plausibly reach the factor of $2-3$ required to bring our simulated data into agreement with the Perseus observations in Figure \ref{fig:linenh}. For the rarer CO isotopologues, the differences between observed and simulated FWHMs become less severe ($2$ versus $1 \kms$ for $^{13}$CO, and $1$ versus $0.7 \kms$ for C$^{18}$O) and the line intensities are in correspondingly better agreement. This again suggests that discrepancies between our simulated cloud and the Perseus data primarily originate in the outer, lower-density regions of the cloud.}

Line emission from molecules with similar chemical behaviour to CO (HCN, CS, HNC; \citealt{priestley2023c}) differs from the Perseus data in our simulated observations. The intensities rise sharply with column density up to $\sim 10^{22} \pcs$ and then saturate at a near-constant value, whereas the observed behaviour is a continuous linear rise between $10^{21-23} \pcs$. For species which preferentially trace denser gas (NH$_3$, N$_2$H$^+$, HCO$^+$), the simulation is in {somewhat} better agreement with the data. As with the CO lines, this suggests that while our simulated cloud differs from Perseus in some respects along moderate-density sightlines ($\sim 10^{21} \pcs$), we are accurately capturing the physical and chemical properties of the densest regions. We note that while the linear relationship between line intensity and column density has also been observed in the Orion A and California clouds \citep{tafalla2023}, similar studies of Orion B \citep{pety2017,santamaria2023} and W49 \citep{barnes2020} find plateau-like behaviour for HCN and other molecules, which more closely resembles our simulated cloud. We discuss interpretations of the unexpected linear relationship for optically-thick lines in Section \ref{sec:linear}.

\subsection{Tracers of dense gas}

Figure \ref{fig:lineratios_img} shows maps of the intensity ratios of the $^{12}$CO, HCN and N$_2$H$^+$ lines seen face-on, and Figure \ref{fig:lineratios_nh} shows how their values vary with column density. Edge-on maps, as with the total intensity maps, are qualitatively similar to their face-on equivalents. The HCN/$^{12}$CO ratio is frequently used as an indicator of the fraction of dense gas (with the exact interpretation of `dense' varying) on extragalactic scales, but in our simulated cloud, its value {varies by only a factor of two} over two orders of magnitude in column density, suggesting it is a poor indicator of genuinely-dense material. Ratios of other lines, such as CS and HCO$^+$, also show relatively little variation over the entire range of column densities where they are detectable (above $3 \times 10^{21} \pcs$). As argued by several observational studies \citep{pety2017,kauffmann2017,tafalla2021}, ratios involving N$_2$H$^+$ are much more sensitive to the presense of material above a column density of $10^{22} \pcs$, thought to be related to the onset of star formation \citep{lada2010}, {although effects not considered in our simulation such as protostellar heating may complicate this interpretation \citep{feher2024}}.

Of particular note, the N$_2$H$^+$/HCN ratio in our cloud rises by a factor of ten over a relatively small range in column density, making it a sensitive probe of high-density material. This line ratio has become observationally accessible on extragalactic scales in recent years, with a typical value of $\sim 0.2$ \citep{jimenez2023,stuber2023}. On the scale of our simulated cloud, this value corresponds closely to the $10^{22} \pcs$ star formation threshold {found by \citet{lada2010}}, suggesting that a significant fraction of the $\kpc$-scale beam area in these extragalactic studies is made up of actively star-forming gas.

The inadequacy of HCN as a dense gas tracer is also apparent in velocity-resolved data. Figure \ref{fig:cloudprofile} shows the face-on line profiles for $^{12}$CO, {the central HCN hyperfine component,} and {the isolated N$_2$H$^+$ component}, averaged over the central $16.2 \pc$ region, as compared to the actual line-of-sight velocity distribution of the mass in this region. Both the $^{12}$CO and HCN lines are quite accurate tracers of the whole-cloud dynamics, and in fact have almost indistinguishable profile shapes. The N$_2$H$^+$ profile is significantly narrower, being a much better (although still imperfect) tracer of the dynamics of the high-density ($>10^4 \pcc$) material.

The difference in the velocity dispersions measured via different tracers has been noted previously in observations \citep{goodman1998,pineda2010,ragan2015,peretto2023}, and appears to represent a real reduction in the magnitude of turbulent motions in the densest material, possibly due to dissipation in accretion shocks \citep{klessen2005,priestley2023b}. This means that the observational definition of velocity dispersion, used to determine quantities such as virial stability \citep[e.g.][]{rigby2024} and magnetic field strength \citep{wang2024}, is tracer-dependent in a highly non-trivial manner. A structure's virial ratio determined via the HCN line will be different to the one calculated using the N$_2$H$^+$ line, and the `correct' value to use will depend entirely on the underlying science question. The association between a line's velocity structure and the underlying gas dynamics should be made with considerable caution.

\subsection{Core-scale line properties}

To investigate the line emission properties on the scales of prestellar cores, we identify nine peaks in the face-on column density map, shown in Figure \ref{fig:coreimg}, and extract line profiles from $0.1 \pc$ apertures around these peaks, approximating the effects of beam size for single-dish observations of nearby clouds \citep[e.g.][]{lee1999,tafalla2002}. Figure \ref{fig:corelines} shows the resulting line profiles of three species commonly used to trace core-scale emission: C$^{18}$O, HCN, and N$_2$H$^+$ {(we again focus on the central HCN and isolated N$_2$H$^+$ hyperfine components)}. The HCN line is in most cases substantially broader than the other two, and often shows evidence of multiple velocity components which are not visible in the other two lines. Rather than selectively tracing the densest `core' material, HCN emission also originates from gas with significantly lower densities ($\ll 10^4 \pcc$; \citealt{evans2020,jones2023}), making it a poor tracer of the cores' dynamics. Several cores also show evidence of multiple velocity components in the N$_2$H$^+$ profiles, suggestive of line-of-sight confusion between physically-distinct regions of dense gas. These features are rarely seen in observational samples of nearby cores with comparable ($\sim 0.1 \pc$) spatial resolution \citep{lee1999,tafalla2002}, although complex multi-component N$_2$H$^+$ profiles on similar spatial scales do occur in more distant infrared-dark clouds \citep{rigby2024}.

Figure \ref{fig:p22} shows the peak line intensity of CS versus that of N$_2$H$^+$ for the simulated cloud core sample, compared to observational data from \citet{lee1999} and \citet{tafalla2002}. The ratio of these lines was identified by \citet{yin2021} as a diagnostic of the degree of magnetic support, based on simulations of isolated, initially-spherical cores. \citet{priestley2022} found that model cores with supercritical mass-to-flux ratios \citep{mouschovias1976}, also shown in Figure \ref{fig:p22}, were incapable of matching the observed distribution of this ratio, as in the absence of magnetic support collapse proceeds too rapidly to significantly deplete CS from the gas phase, and its $J=2-1$ line is subsequently stronger than observed. By contrast, the cores formed in our simulation agree well with the observational data, despite the initial mass-to-flux ratio of the simulation being $2.4$ times the critical value (i.e. moderately supercritical). The complex formation histories of the cores in our simulation, involving material being repeatedly cycled in and out of high-density phases before undergoing runaway gravitational collapse \citep{priestley2023c}, results in them having very different chemical properties to equivalent models of isolated cores, which instead experience a monotonic increase in density with time.

Another problematic aspect of supercritical core models is their predicted line widths, which are inevitably much broader than the observed subsonic values due to the supersonic infall velocities of unimpeded gravitational collape \citep{larson1969}. Figure \ref{fig:p22} shows the FWHMs\footnote{{In all cases measured directly from the line profiles, rather than being the FWHM of a Gaussian fit to the lines.}} of the CS and N$_2$H$^+$ lines of the cores from our simulation and from the \citet{priestley2022} supercritical sample, compared to observational data from \citet{tafalla2002}. In this case, forming cores self-consistently does not improve the agreement with observations: simulated cores still have much broader lines ($> 0.5 \kms$) than real ones. Previous studies \citep{offner2008,smith2012,priestley2023b} have reproduced subsonic line widths in cores and filaments formed via supersonic converging flows, which efficiently dissipate the initial kinetic energy \citep{klessen2005}, so the discrepancy here may be due to our initial turbulent velocity field being somewhat subsonic, even though the cloud collision itself is highly supersonic. Alternatively, simulations may still require a substantial degree of magnetic support to reproduce the observed properties of real objects, but it is clearly impossible to make this argument based on models of isolated spheres, which are far too idealised compared to the complex formation histories of cores in turbulent molecular clouds.

\section{Discussion}

\begin{figure*}
  \centering
  \includegraphics[width=\columnwidth]{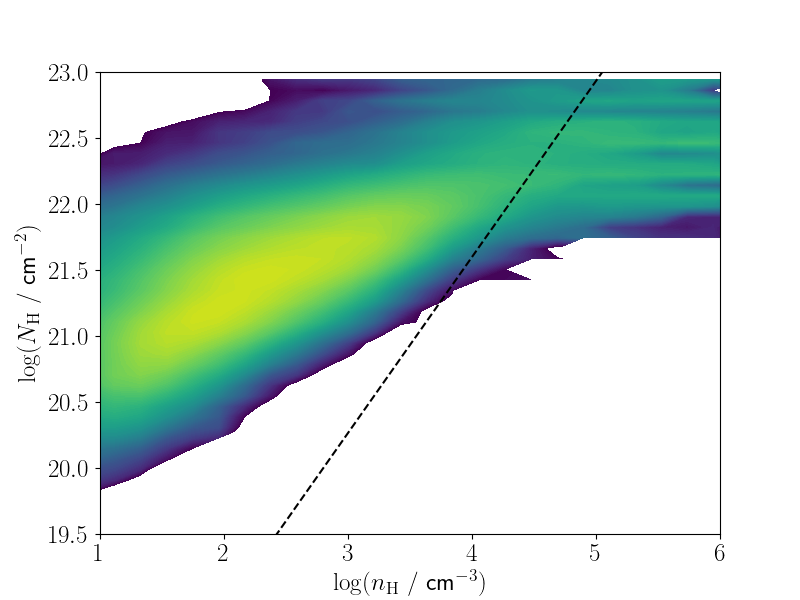}
  \includegraphics[width=\columnwidth]{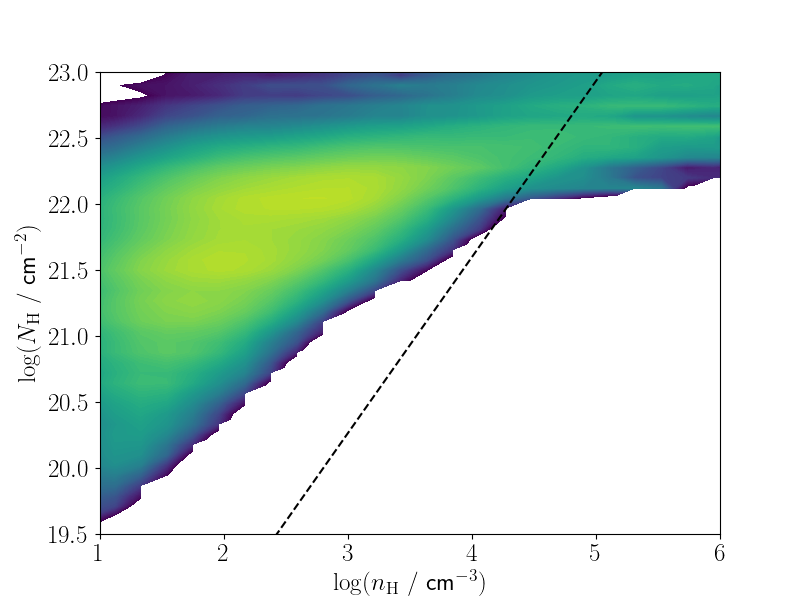}
  \caption{Distribution of {\sc arepo} cells by volume density and column density as seen face-on (left) and edge-on (right). The dashed black lines show the relationship proposed by \citet{tafalla2021} to explain linear correlations between line intensities and column density.}
  \label{fig:nhnh}
\end{figure*}

\begin{figure*}
  \centering
  \includegraphics[width=\columnwidth]{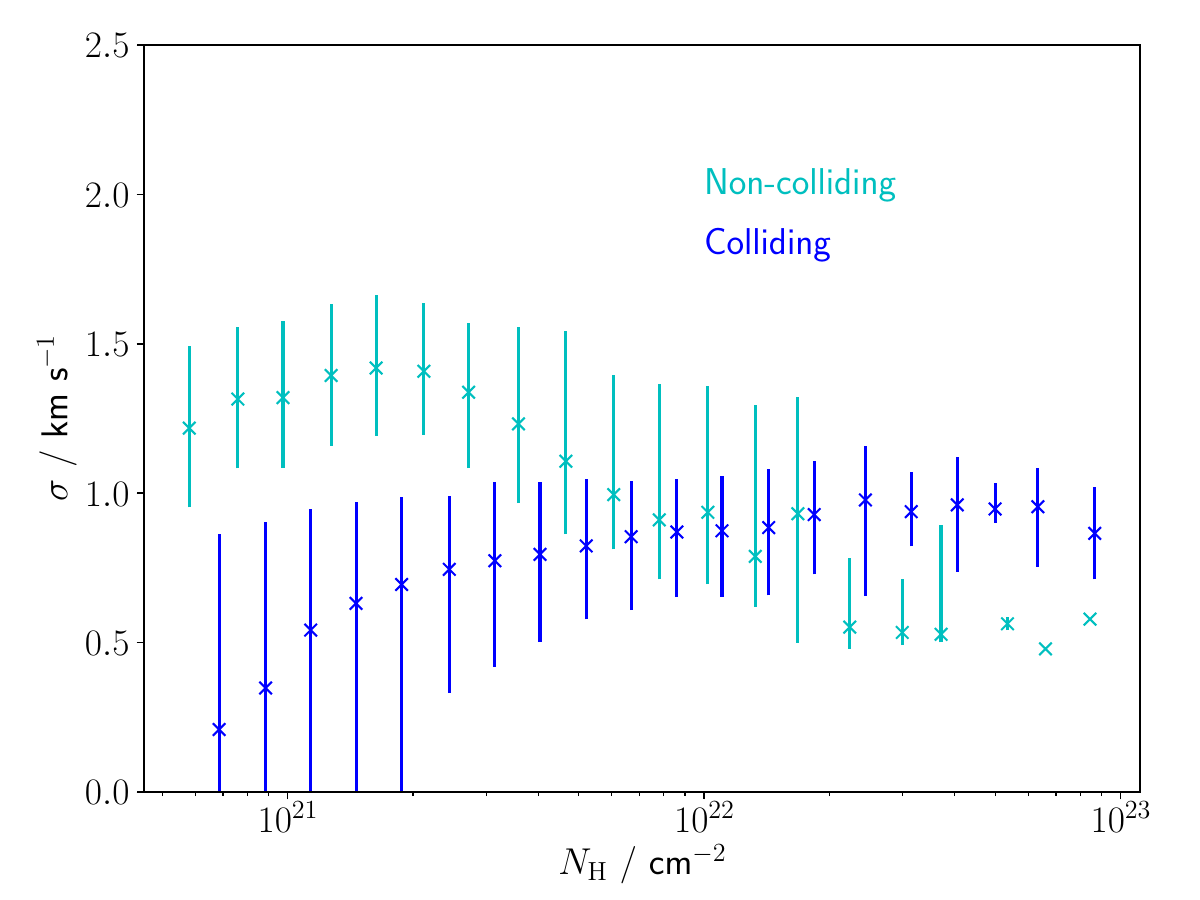}
  \includegraphics[width=\columnwidth]{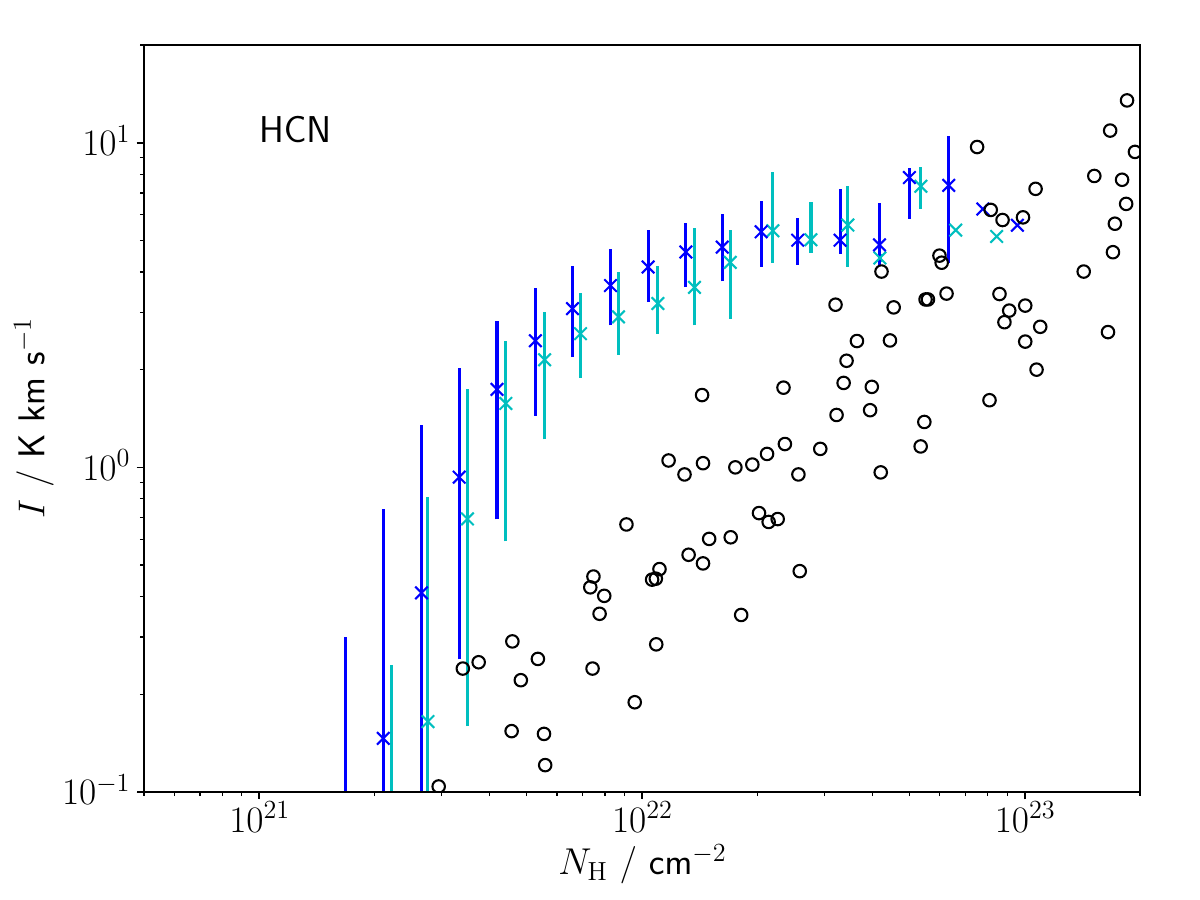}\\
  \caption{Line-of-sight velocity dispersion (left) and HCN line intensity (right) versus column density for the colliding (blue) and non-colliding (cyan) simulations. Crosses show the median pixel values, error bars the 16th/84th percentiles. Observational data from the Perseus molecular cloud for HCN \citep{tafalla2021} are shown in black.}
  \label{fig:sigv}
\end{figure*}

\subsection{The (in)accuracy of isolated core models}

Simulations of isolated prestellar cores are widely used to interpret molecular line observations \citep{keto2015,sipila2018,young2019,sipila2022,tritsis2023,redaelli2024}, under the assumption that cores are sufficiently decoupled from their environment to evolve more-or-less independently. This does not appear to be the case; \citet{wurster2024} find that resimulating the regions which form cores within a larger-scale cloud produces significantly different results to the original simulation, due to the chaotic nature of star formation, while several observational studies \citep{peretto2020,rigby2021,anderson2021,redaelli2022} have suggested that cores continue to accrete mass from their surroundings throughout their lifetimes.

The prior evolutionary history of the material making up a core, and continuing accretion onto it, can significantly alter its chemical composition \citep{priestley2023c}, with a correspondingly large effect on the predicted line emission properties. The peak line intensities of our core samples in Figure \ref{fig:p22} resemble those of magnetically-supported isolated models from \citet{priestley2022}, despite the opposite being the case in the simulation, while the subsonic linewidths of observed cores may be a result of their formation from cloud-scale turbulent motions \citep{offner2008,smith2012,priestley2023b} rather than the isolated quasi-equilibrium scenario adopted by \citet{keto2015} and subsequent other authors. Taking these highly-idealised models as representative of real cores risks making fundamental errors in the physical interpretation of molecular line data.

We note that a similar argument could be made regarding our idealised setup of two colliding, uniform-density spheres. Real molecular clouds are rarely spherical \citep{faustino2024}, and may be better characterised by a continuous injection of turbulence \citep{zhou2022}, rather than the decaying velocity field we employ. However, the generally-good agreement between the predicted line emission from our simulations and observational data, on both cloud and core scales, suggests that they are accurately capturing the physical-chemical structure of real molecular clouds (see also \citealt{priestley2023a}), despite their idealised initial conditions. Simulations on the galactic disc scales necessary to self-consistently follow cloud formation \citep[e.g.][]{panessa2023} typically lack the resolution to investigate chemistry in detail in pre- and protostellar cores. These objects are likely to be sufficiently decoupled from their galactic environment for cloud-scale simulations to be an entirely adequate representation of their formation and evolution.

\subsection{Sulphur depletion in molecular gas}
\label{sec:sulphur}

It is common practice in astrochemical modelling to reduce the elemental gas-phase sulphur abundance by a factor of ten or more from its Solar value, as otherwise many sulphur-bearing molecules are predicted to be far more abundant, and their line emission far stronger, than is actually observed \citep{fuente2019,navarro2020,fuente2023}. The assumption is that most of the sulphur is locked up in some chemically-unavailable solid phase, but direct measurements of sulphur depletion in the interstellar medium find a near-Solar gas-phase abundance even along the densest sightlines \citep{jenkins2009}, suggesting that only a modest fraction of the total sulphur is in solid form.

Our simulation, using an undepleted sulphur abundance (Table \ref{tab:abun}), successfully reproduces the observed peak intensities of CS in cores (Figure \ref{fig:p22}), often one of the more problematic species in astrochemical modelling \citep{navarro2020}. As argued by \citet{hily2022}, the main sulphur reservoir in molecular clouds may be atomic S, rather than sulphur-bearing molecules \citep{holdship2019,kushwahaa2023} and ices \citep{boogert2015,mcclure2023}, which make up only a few percent of the Solar abundance when combined. It was speculated in \citet{priestley2022} that the sulphur abundance problem may be resolved by more realistic physical models of molecular gas, rather than an improved understanding of its chemistry. We will investigate this possibility, and the detailed behaviour of sulphur-bearing species in our chemical model, in future work.

\subsection{Linear intensity scaling in optically-thick lines}
\label{sec:linear}

The linear relationship between line intensity and column density for molecules such as HCN in Perseus \citep{tafalla2021} and some other molecular clouds \citep{tafalla2023} is unexpected, as these lines are optically thick {(typical line-centre opacities $>10$ for column densities $> 10^{21} \pcs$)}. The amount of emitting material along the line-of-sight should then have little impact on the strength of the observed emission. Explanations for this behaviour \citep{tafalla2021,dame2023} typically rely on the presence of a positive correlation between volume and column densities; if the characteristic volume density is larger along higher-column sightlines, and the level populations are subthermally-excited, then an increasing column density can produce an increase in line intensity even in the optically-thick limit.

Simulated molecular clouds typically do show such a relationship \citep{clark2014,bisbas2019}, but not one capable of reproducing the linear behaviour observed by \citet{tafalla2021}. Figure \ref{fig:nhnh} shows the distribution of the volume and line-of-sight column densities of {\sc arepo} cells in our simulation. The correlation between column and volume density is much flatter than the $\Nhcol \propto \nh^{4/3}$ proposed by \citet{tafalla2021}, and more significantly, departs severely from a one-to-one relationship; a given line-of-sight column density can comprise gas covering several orders of magnitude in volume density. It seems probable that at least the latter behaviour is shared with real molecular clouds, making the excitation temperature argument for the linear intensity scaling unviable; there should be little, if any, correlation between the excitation temperature of a transition, a complicated mass- and optical depth-weighted average over material of all densities along the line of sight, and the observed column density. {We note that despite the face-on and edge-on cloud orientations having very different volume density distributions for line-of-sight columns above $10^{22} \pcs$, the resulting line intensities in this regime in Figure \ref{fig:linenh} are effectively indistinguishable, suggesting that the characteristic volume density of a line of sight has a negligible effect on the resulting line emission.}

An alternative explanation for the increasing intensity of optically-thick lines is if the line-of-sight velocity dispersion is higher along high-density sightlines, providing more `bandwidth' available for emission \citep{whitworth2018}. Simulations in \citet{priestley2023a} where the velocity dispersion increases with column density produced linearly-increasing HCN line intensities, whereas simulations without a rise in velocity dispersion resulted in an intensity plateau similar to those in Figure \ref{fig:linenh}. Our simulation has an almost-constant velocity dispersion over the range of column densities where HCN is detectable (Figure \ref{fig:sigv}), so no additional velocity channels become available for emission, and once the HCN line becomes optically thick, its intensity saturates.

The distinction between rising and flat velocity dispersions in \citet{priestley2023a} was attributed to simulations of isolated versus colliding clouds. However, we do not reproduce this behaviour here. Figure \ref{fig:sigv} shows the velocity dispersion and HCN intensity relationships for a simulation where we have reduced the collision velocity from $7$ to $0.5 \kms$; the subsonic collision velocity (also below the average velocity of the initial turbulent field) results in behaviour comparable to an isolated, turbulent cloud as studied in \citet{priestley2023a}. Rather than increasing, the velocity dispersion actually declines over the range of column densities relevant for HCN emission, resulting in line intensities which are indistinguishable from the colliding case. We attribute this to our use of a lower initial density compared to \citet{priestley2023a} ($10$ versus $250 \pcc$), so that the initial turbulent velocity field is dissipated before the cloud material has had time to collapse to the higher densities where HCN emission can be produced. Establishing the relevance of this distinction to the interpretation of observational data is left to future work.

\section{Conclusions}

We have combined an MHD simulation of {the formation and dynamical evolution of} a molecular cloud with time-dependent chemistry and radiative transfer modelling, producing self-consistent synthetic observations in {the $J=1-0$ rotational transitions of $^{12}$CO, $^{13}$CO, C$^{18}$O, HCN, N$_2$H$^+$, HCO$^+$ and HNC, the $J=2-1$ transition of CS, and the $(1,1)$ inversion transition of para-NH$_3$}. Our main results are as follows:
\begin{itemize}
\item
  Our simulated cloud reproduces the observed range of intensities for most lines well, with {better} agreement for species which preferentially trace high-density regions. This suggests that despite the idealised initial conditions and modelling assumptions, the simulation has produced a molecular cloud with similar structural properties to real ones.
\item
  HCN, along with many other molecules often described as dense-gas tracers, is actually a poor tracer of dense gas, being detectable down to low column density. {The velocity information derived from the HCN line profile mostly traces the bulk dynamics of the cloud.} N$_2$H$^+$ is a much more selective tracer of high-density ($>10^{22} \pcs$) regions, and its line profile shape is strongly correlated with the dynamics of material above $10^4 \pcc$.
\item
  Cores formed from the large-scale cloud dynamics in the simulation agree well with the distribution of line intensities from observational samples, although their line widths are somewhat broader. They do not agree with the predictions of an equivalent sample of isolated core models, which neglect the formation of the cores themselves from their more diffuse environment. Using these isolated core models to interpret molecular line observations is likely to result in erroneous conclusions.
\item
  {We obtain CS line intensities in good agreement with observed values using an undepleted elemental abundance of sulphur. This suggests that the common modelling practice of reducing this quantity by several orders of magnitude to match observations of sulphur-bearing species may be a sign of the limitations of the physical models employed, rather than genuine evidence for sulphur depletion in molecular clouds.}
\end{itemize}
Our work demonstrates the importance of simulating the large-scale dynamics of molecular clouds when investigating their line emission properties, even when the subject of interest is on much smaller scales. This serves as a starting point for future work to fully exploit the ability of molecular line emission to advance our understanding of star formation.

\section*{Acknowledgements}

We are grateful to Rachel Friesen for providing the GAS data, and to Nicolas Peretto for a useful discussion about the comparison with observations. FDP, PCC, SER and OF acknowledge the support of a consolidated grant (ST/K00926/1) from the UK Science and Technology Facilities Council (STFC). SCOG and RSK acknowledge funding from the European Research Council (ERC) via the ERC Synergy Grant “ECOGAL-Understanding our Galactic ecosystem: From the disk of the Milky Way to the formation sites of stars and planets” (project ID 855130), from the Heidelberg Cluster of Excellence (EXC 2181 - 390900948) “STRUCTURES: A unifying approach to emergent phenomena in the physical world, mathematics, and complex data”, funded by the German Excellence Strategy, and from the German Ministry for Economic Affairs and Climate Action in project ``MAINN'' (funding ID 50OO2206). The team in Heidelberg also thanks for computing resources provided by {\em The L\"{a}nd} through bwHPC and DFG through grant INST 35/1134-1 FUGG and for data storage at SDS@hd through grant INST 35/1314-1 FUGG. LRP acknowledges support from the Irish Research Council Laureate programme under grant number IRCLA/2022/1165. This research was undertaken using the supercomputing facilities at Cardiff University operated by Advanced Research Computing at Cardiff (ARCCA) on behalf of the Cardiff Supercomputing Facility and the Supercomputing Wales (SCW) project. We acknowledge the support of the latter, which is part-funded by the European Regional Development Fund (ERDF) via the Welsh Government. This research has made use of spectroscopic and collisional data from the EMAA database (https://emaa.osug.fr and https://dx.doi.org/10.17178/EMAA). EMAA is supported by the Observatoire des Sciences de l’Univers de Grenoble (OSUG)

\section*{Data Availability}
The data underlying this article will be shared on request.

\bibliographystyle{mnras}
\bibliography{survey}

\appendix

\section{Resolution tests}
\label{sec:restest}

\begin{figure}
  \centering
  \includegraphics[width=\columnwidth]{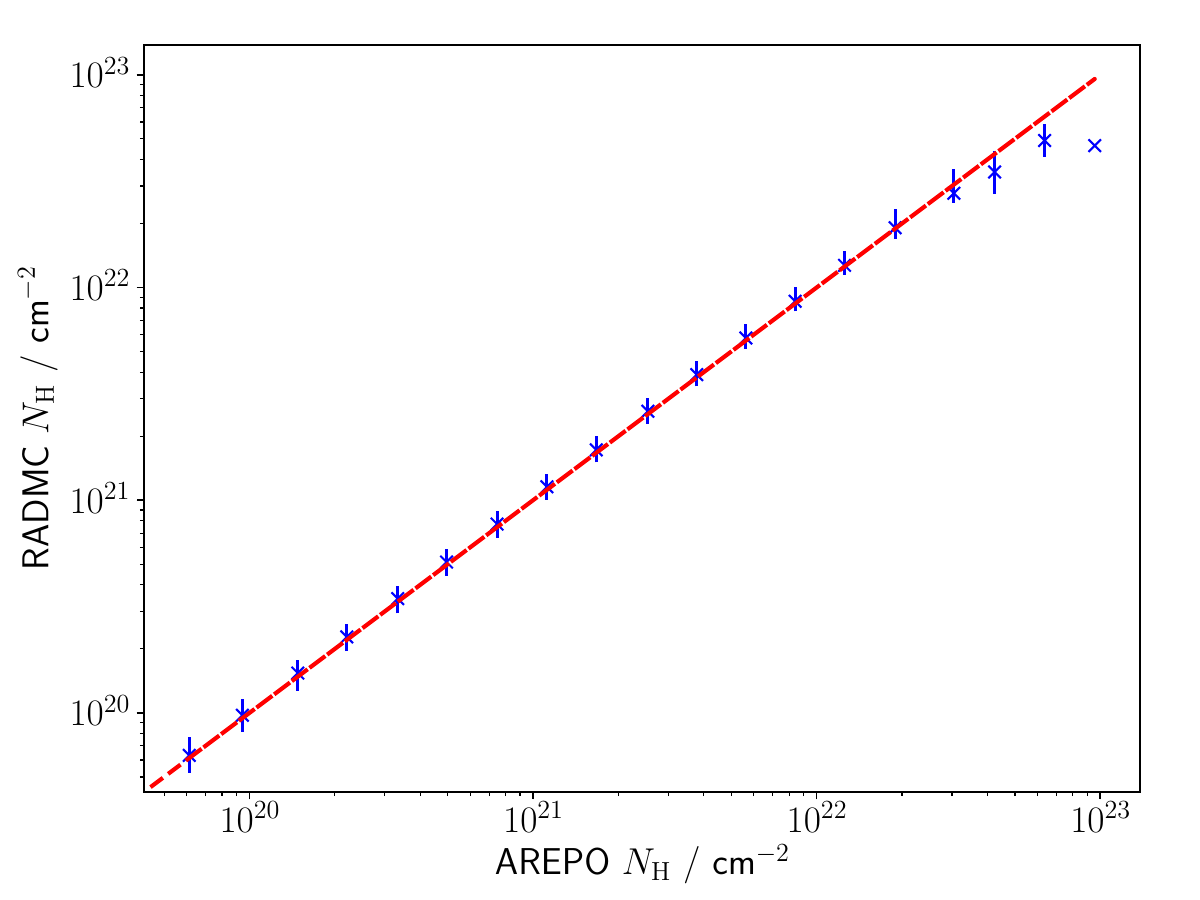}
  \caption{Column density of the {\sc radmc3d} grid versus the original {\sc arepo} Voronoi mesh. Crosses show the median pixel values, error bars the 16th/84th percentiles. The dashed red line shows the expected one-to-one relationship.}
  \label{fig:resdust}
\end{figure}

\begin{figure*}
  \centering
  \includegraphics[width=0.32\textwidth]{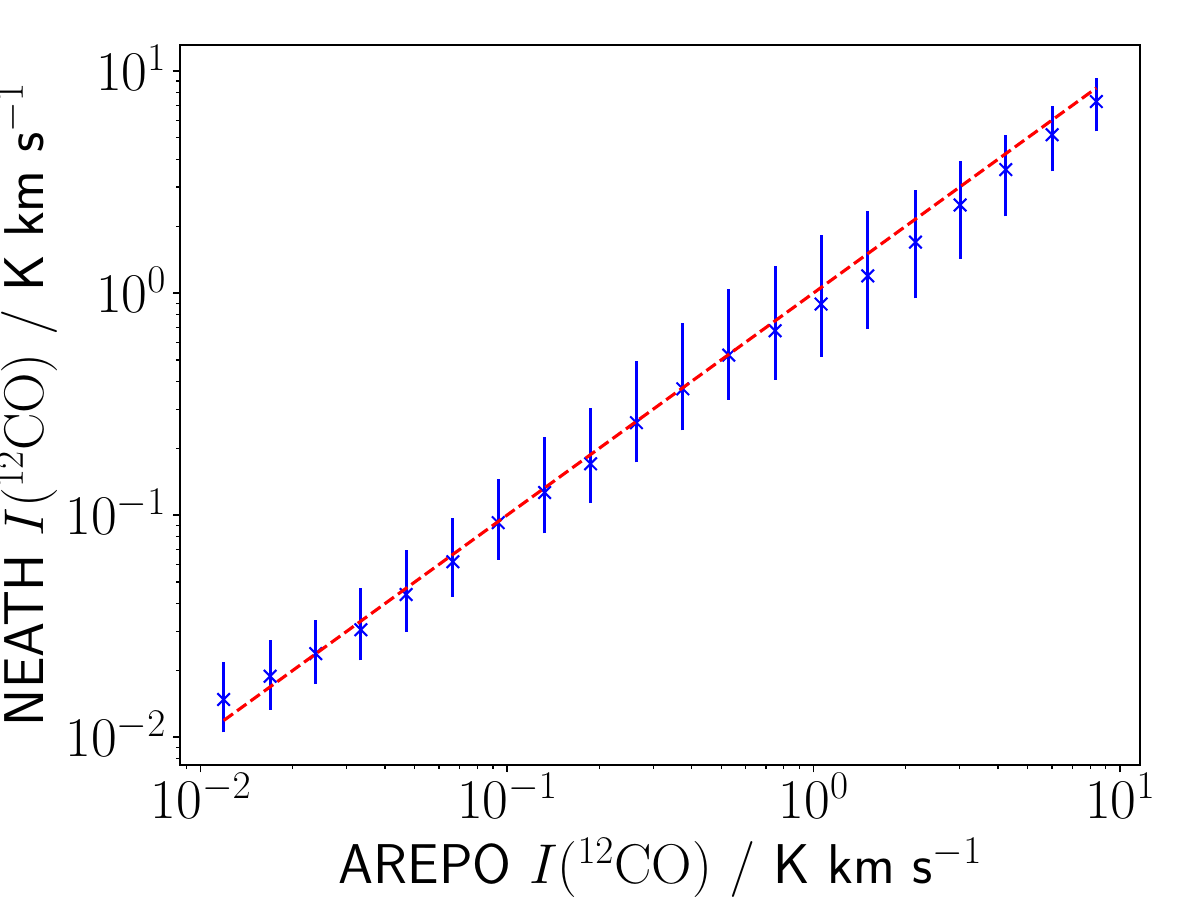}
  \includegraphics[width=0.32\textwidth]{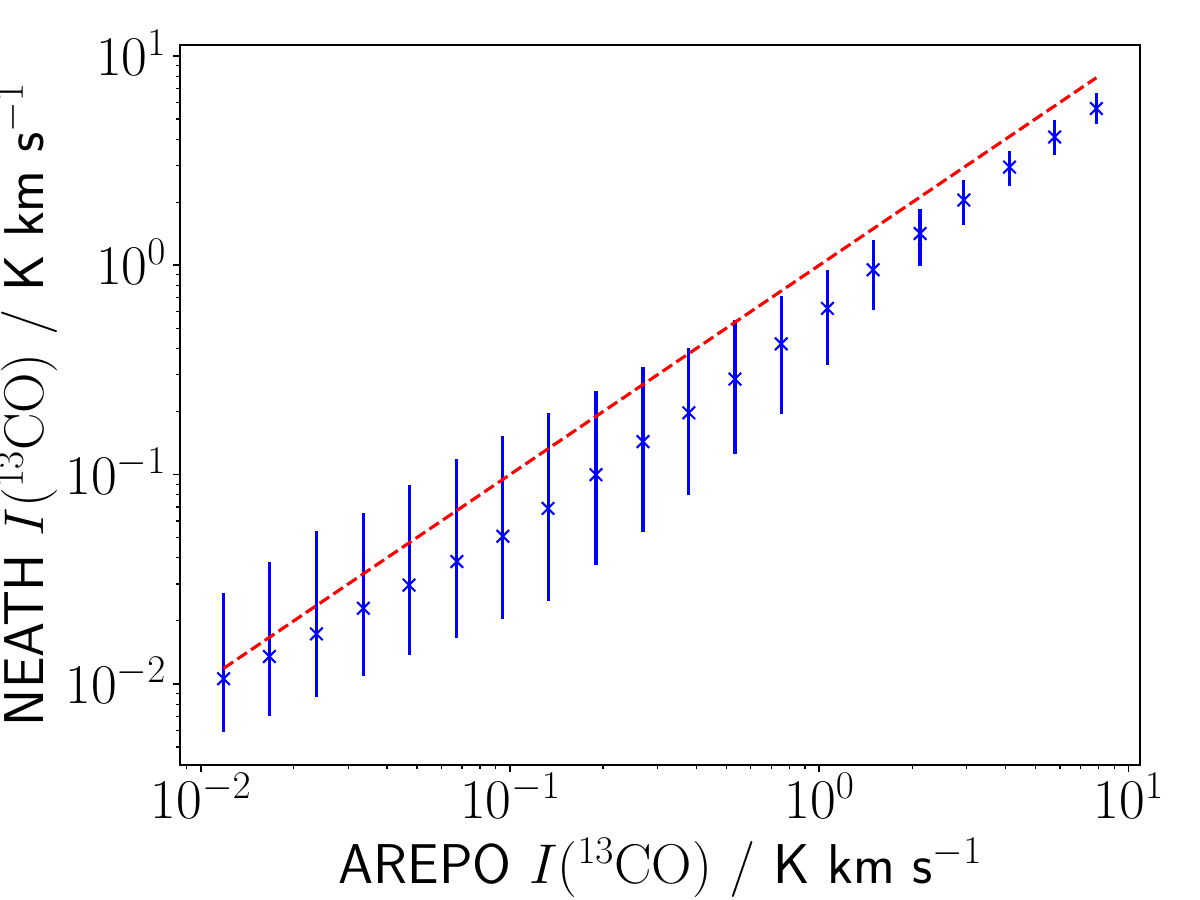}
  \includegraphics[width=0.32\textwidth]{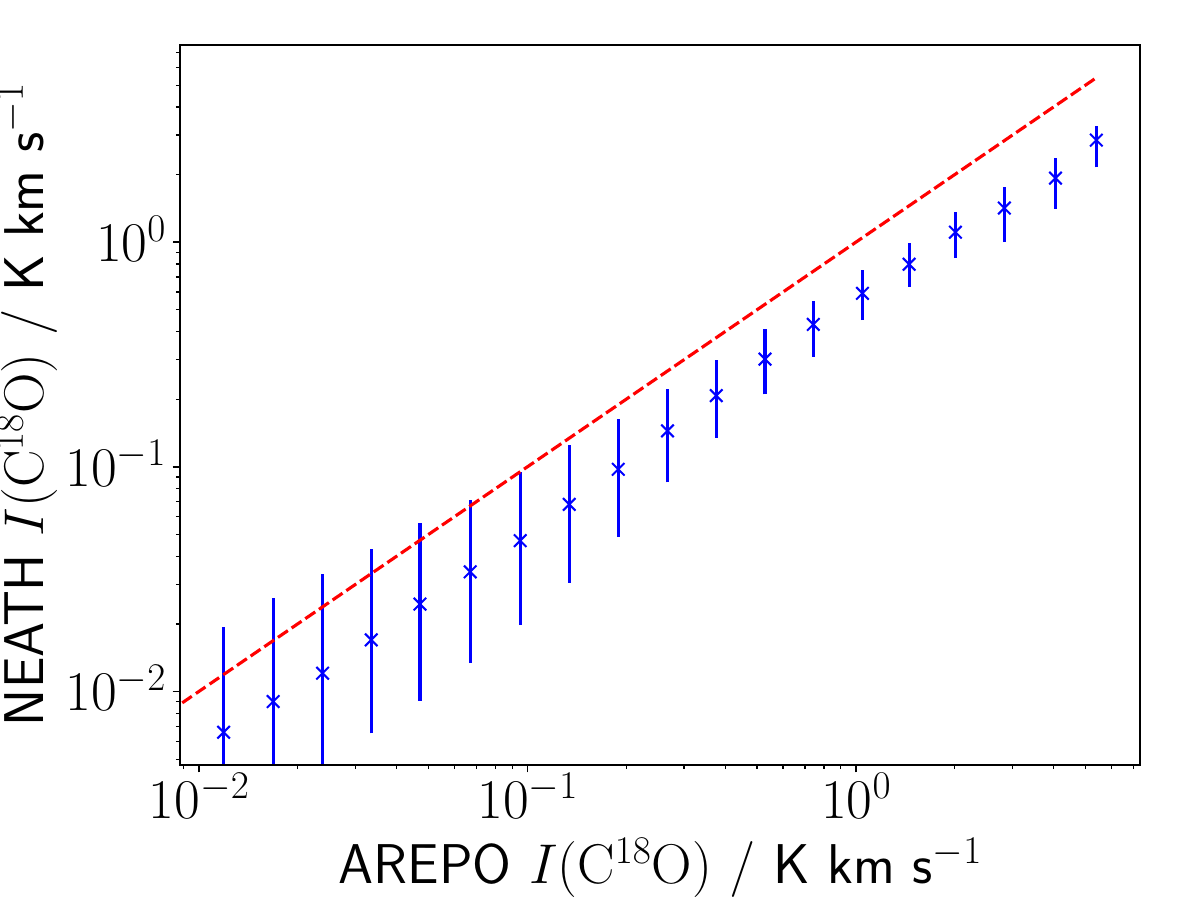}
  \caption{Integrated $^{12}$CO (left), $^{13}$CO (centre) and C$^{18}$O (right) intensities using the NEATH tracer particle abundances versus the AREPO cell values. Crosses show the median pixel values, error bars the 16th/84th percentiles. The dashed red lines shows the expected one-to-one relationship.}
  \label{fig:resco}
\end{figure*}

The MHD, chemical and radiative transfer simulations in this paper each have mutually-incompatible concepts of resolution, which results in some losses when properties are mapped between them. The cubic adaptive mesh used by {\sc radmc3d} loses detail when compared to the unstructured Voronoi mesh used by {\sc arepo}, and with only $10^5$ post-processed tracer particles compared to several million grid cells, the information on molecular abundances is inevitably `smeared out' to some extent. We demonstrate below that neither of these effects is large enough to substantially alter our results.

Figure \ref{fig:resdust} shows the relationship between the true column density obtained from the {\sc arepo} Voronoi mesh, and the column density from the {\sc radmc3d} grid used for the radiative transfer. We determine the latter by fixing the dust temperature to $10 \kel$ in all cells, and producing an image of the cloud at $850 \um$, where the dust optical depths are $\lesssim 10^{-4}$. As the dust opacity and gas-to-dust ratio are both constant, the intensity is directly proportional to the column density of the {\sc radmc3d} grid. After converting from intensity to $\Nhcol$, the {\sc radmc3d} column densities are almost identical to the true values, with the exception of the highest end of the range near $10^{23} \pcs$. Even here, the difference is less than a factor of two, and these column densities represent a small fraction of the cloud area, so we consider the mapping between the {\sc arepo} and {\sc radmc3d} grids to be robust.

While the {\sc radmc3d} grid appears to accurately represent the underlying structure of the MHD simulation, the molecular abundances used in the radiative transfer are taken from $10^5$ post-proessed tracer particles, compared to $\sim 20$ million cells in the {\sc radmc3d} grid. We assess the accuracy of this approach using CO emission lines, as the CO abundance for every cell can be taken directly from the {\sc arepo} simulation as a point of comparison to the NEATH tracer particle abundance.

Figure \ref{fig:resco} shows the relationship between integrated intensity of the three CO isopologues for the two abundance determinations. All three lines show a linear correlation in intensity between the NEATH and {\sc arepo} values, albeit with some scatter. The two rarer isotopologues tend to have somewhat lower line intensities for the NEATH abundances than when using the {\sc arepo} values, but this can be attributed to the neglect of freeze-out in the {\sc arepo} chemical model, which severely reduces the CO abundance at high volume densities \citep{priestley2023c}. This effect is not visible for the more-abundant $^{12}$CO line, which is too optically-thick to probe the densities where freeze-out occurs. The otherwise-good correlation for all isotopologues suggests that the number of tracer particles is adequate to correctly capture the molecular structure of the cloud.

\section{Dust-derived column densities}
\label{sec:dust}

\begin{figure*}
  \centering
  \includegraphics[width=\columnwidth]{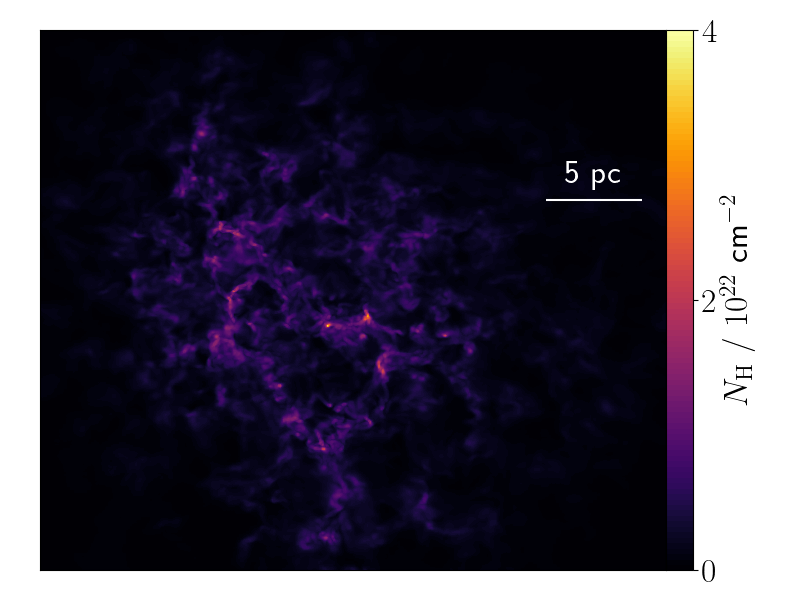}
  \includegraphics[width=\columnwidth]{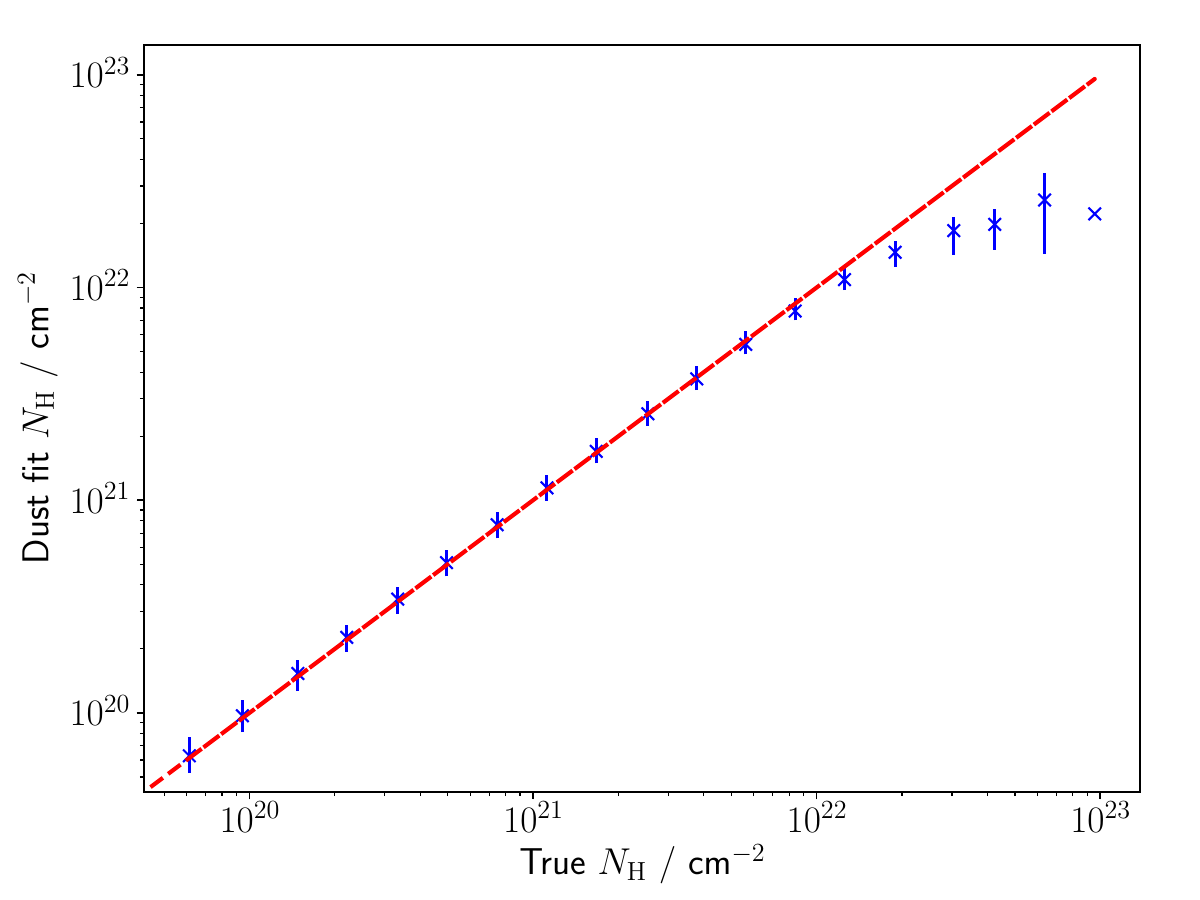}
  \caption{{\it Left:} Column density map of the cloud derived from the far-IR dust emission (compare Figure \ref{fig:coldens}). {\it Right:} True column density versus dust-derived column density. Crosses show the median pixel values, error bars the 16th/84th percentiles. The dashed red line shows the expected one-to-one relationship.}
  \label{fig:dustcol}
\end{figure*}

 Figure \ref{fig:dustcol} shows the column density of the cloud inferred from thermal dust emission, the most common means of assessing molecular cloud structures observationally. We produce images of the cloud at wavelengths of $100$, $160$, $250$, $350$ and $500 \um$, corresponding to the central wavelengths of the {\it Herschel} filters generally used to study cold molecular gas \citep[e.g.][]{ragan2012b,konyves2015}. We fit the resulting pixel-by-pixel spectral energy distributions (SEDs) with a single-temperature modified blackbody, using the same dust opacity as in the radiative transfer calculation, and the same dust-to-gas ratio to convert dust mass into total column density. Any differences between the two column densities are therefore due to temperature variations along the line of sight \citep{shetty2009}.

 The dust-derived column densities are generally in good agreement with the true values. However, this breaks down at the highest values, with peaks in the true column density images less prominent in the dust-column maps. The correlation between the column densities derived from dust emission and the true values, shown in Figure \ref{fig:dustcol}, are nearly identical up to true columns of $10^{22} \pcs$. Beyond this point, the column densities derived from dust emission are lower than the true values by factors of up to $2-3$, as an increasing proportion of the mass is made up of colder, less emissive grains, whereas the emission is still dominated by the highest grain temperatures. The luminosity-weighted temperature of the blackbody fit is therefore higher than the mass-weighted temperature needed to retrieve the physical column density, which is correspondingly underestimated. As we are mainly concerned with the line emission properties of the cloud, we use the true simulation column densities for simplicity, but note that this effect could become important for the highest-density sightlines.

% Don't change these lines
\bsp	% typesetting comment
\label{lastpage}
\end{document}